\documentclass[%
 reprint,
 superscriptaddress,
nofootinbib,
 amsmath,amssymb,
 aps,
longbibliography,
prx
]{revtex4-1}

\usepackage{graphicx}
\usepackage{soul}
\usepackage{float}
\usepackage{latexsym,amsmath,amssymb,bm,euscript,multirow,makecell}
\usepackage{color,esvect}
\usepackage{wasysym}
\usepackage{epstopdf}
\usepackage[pdftex,colorlinks=true,linkcolor=darkblue,citecolor=blue,urlcolor=darkred]{hyperref}
\usepackage{type1cm}
\usepackage{appendix} 
\usepackage{mathtools}

\usepackage{xcolor}
\definecolor{darkblue}{rgb}{0.,0.,0.4}
\definecolor{darkred}{rgb}{0.5,0.,0.}
\definecolor{BlueViolet}{RGB}{138,43,226}
\definecolor{SkyBlue}{RGB}{30,144,255}
\definecolor{DarkGreen}{RGB}{0,100,0}

\usepackage[normalem]{ulem}
\usepackage{comment}

\newcommand{\tj}[6]{ \begin{pmatrix}
   #1 & #2 & #3 \\
   #4 & #5 & #6 
  \end{pmatrix}}

\renewcommand{\vec}[1]{\bm{#1}}

\begin{document}

\title{Uncovering conformal symmetry in the $3D$ Ising transition: \\ State-operator correspondence from a fuzzy sphere regularization}

\author{Wei Zhu}
\email{zhuwei@westlake.edu.cn}
\affiliation{School of Science, Westlake University, Hangzhou, 310030, China}

\author{Chao Han}
\affiliation{Westlake Institute of Advanced Study, Westlake University, Hangzhou, 310024, China}

\author{Emilie Huffman}
\affiliation{Perimeter Institute for Theoretical Physics, Waterloo, Ontario N2L 2Y5, Canada}

\author{Johannes S. Hofmann}
\affiliation{Department of Condensed Matter Physics, Weizmann Institute of Science, Rehovot, 76100, Israel}

\author{Yin-Chen He}
\email{yhe@perimeterinstitute.ca}
\affiliation{Perimeter Institute for Theoretical Physics, Waterloo, Ontario N2L 2Y5, Canada}

\begin{abstract}

The $3D$ Ising transition, the most celebrated and unsolved critical phenomenon in nature, has long been conjectured to have emergent conformal symmetry, similar to the case of the $2D$ Ising transition. 
Yet, the emergence of conformal invariance in the $3D$ Ising transition  has rarely been explored directly, mainly due to unavoidable mathematical or conceptual obstructions. 
Here, we design an innovative way to study the quantum version of the $3D$ Ising phase transition on spherical geometry, using the ``fuzzy (non-commutative) sphere" regularization. 
We accurately calculate and analyze the energy spectra at the transition, and explicitly demonstrate the state-operator correspondence (i.e. radial quantization), a fingerprint of conformal field theory.
In particular, we have identified 13 parity-even primary operators within a high accuracy and 2 parity-odd operators that were not known before.
Our result directly elucidates the emergent conformal symmetry of the $3D$ Ising transition, a conjecture made by Polyakov half a century ago.
More importantly, our approach opens a new avenue for studying $3D$ CFTs by making use of the state-operator correspondence and spherical geometry.

\end{abstract}

\maketitle

%\tableofcontents

\section{Introduction}

Symmetry is one of the most important organizing principles in physics. 
As is well known, symmetries present microscopically (e.g. condensed matter systems, ultraviolet (UV) Lagrangians) can be spontaneously broken at low energies, giving rise to various distinct phases of matter such as crystals and magnets. 
Conversely and rather unexpectedly,  symmetries absent microscopically can emerge at low energies, and such a phenomenon is called emergent symmetry. 
One prominent example is the order-disorder phase transition of $2D$ Ising model, for which Polyakov discovered emergent conformal symmetry in 1970~\cite{polyakov1970conformal}, 26 years after Onsager's exact solution~\cite{Onsager1944}. 

Polyakov's remarkable discovery of emergent conformal symmetry in the $2D$ Ising transition gave birth to conformal field theory (CFT) \cite{Belavin1984}, a class of quantum field theories with profound applications in various fields of physics including statistical mechanics, quantum condensed matter, string theory and quantum gravity. 
In statistical physics, it is a common belief that many universality classes of (classical and quantum) phase transitions are captured by CFTs, however this has not been proven for $3D$ transitions.~\footnote{For phase transitions in $2D$~\cite{polchinski1988scale} and $4D$~\cite{Dymarsky2015scale} the combination of scale symmetry, Lorentz symmetry and unitarity was shown to lead to conformal symmetry.}
The emergence of conformal symmetry at phase transitions is not only aesthetically beautiful, but also useful in understanding the properties of these transitions, such as computing experimentally measurable critical exponents.
In $2D$ the (local) conformal symmetry has an infinite-dimensional algebra, and it makes many $2D$ CFTs exactly solvable \cite{yellowbook,Belavin1984}. 
In $d>2$ 
dimensions, there is only a finite-dimensional (global) conformal symmetry, i.e. $SO(d+1,1)$, with which one is not able to analytically solve CFTs as in $2D$.
Therefore, CFTs beyond $2D$ are rather poorly understood, with their solutions remaining outstanding for decades despite their broad appeal to physics and mathematics. 

Historically, the study of lattice models for $2D$ classical  phase transitions and their quantum cousins ($1+1D$ quantum phase transitions) played a key role in the discovery and understanding of $2D$ CFTs~\cite{Onsager1944,polyakov1970conformal,Cardy1984}.   
Similar progress in the study of conformal symmetry for $d\ge 3$ dimensional theories, however, has stalled due to the natural limitation of the lattice formulation.
There are a plenty of papers studying $3D$ phase transitions on the lattice, e.g. computing critical exponents. However, the perspective of conformal symmetry has rarely been explored~\cite{Weigel2000,Deng2002Conformal,Billo2013Line,Cosme2015Sphere,Schuler2016Universal,Meneses2019viral}.
The conformal symmetry of a $d-$dimensional CFT is most transparent in geometries such as $\mathbb R^d$, $S^d$ as well as  $S^{d-1}\times \mathbb{R}$. 
In particular, CFTs on $S^{d-1}\times \mathbb{R}$ obey a property called state-operator correspondence (i.e. radial quantization), which is a direct consequence of conformal symmetry \cite{Cardy1984}.
Specifically, for a quantum Hamiltonian defined on sphere $S^{d-1}$, its eigenstates are in one-to-one correspondence with the scaling operators (including primary and descendant operators) of the infrared (IR) CFT. 
Moreover, the energy gaps of these eigenstates are proportional to the scaling dimensions of their corresponding scaling operators  \cite{CARDY1985}.
This nice feature can be used to explore various properties of CFTs, including scaling dimensions of operators, operator product expansion coefficients, and even operator algebras~\cite{Cardy1984}.
For $2D$ CFTs, $S^1\times \mathbb{R}$ is very natural as one just needs to study a $1+1D$ quantum lattice model defined on a $1D$ periodic chain (i.e. $S^1$)~\cite{Blote1986Conformal,affleck1988universal,Milsted2017,Zou2018}. 
However, simulating lattice models of $d\ge 3$ dimensional CFTs on $S^{d-1}\times \mathbb{R}$ will be problematic, because a regular lattice cannot be put on a sphere $S^{d-1\ge 2}$ due to its nontrivial curvature.~\footnote{In mathematics the problem of tiling a sphere is called spherical tiling or spherical polyhedron.}
While efforts have nevertheless been made to discretize the sphere, no signature of state-operator correspondence has been found so far~\cite{Brower2013Lattice,Brower2021Radial}. 

To overcome this geometric obstacle, in this paper we are pursuing a different direction, namely we fuzzify a sphere~\cite{madore1992fuzzy}. 
Specifically, we study a $2+1D$ quantum Ising transition defined on a fuzzy (non-commutative) sphere in light of Landau level regularization~\cite{Ippoliti2018Half}. 
As a result of this innovative discretization, we have observed almost perfect state-operator correspondence in surprisingly small system sizes. 
We use exact diagonalization to calculate properties of the $2+1D$ Ising transition for up to $16$ effective spins, and we have found its low lying eigenstates (up to 70 lowest states) split into representations of the $3D$ conformal symmetry (i.e. conformal multiplets), hence directly demonstrating the emergence of conformal symmetry.
Among these low energy states, we have found 15 conformal primary states, 
most of which have not been discovered in any previous model
studies of the 3D Ising transition.
Specifically, we have found 13 parity-even primaries, whose scaling dimensions agree well with  state-of-the-art conformal bootstrap results~\cite{RMP_CB,Ising_CB} with discrepencies smaller than $1.6\%$.
We have also identified two parity-odd primaries which were
unknown before.

Our observations directly verify conformal symmetry for the $3D$ Ising transition, which was conjectured by Polyakov 50 years ago \cite{polyakov1970conformal}. 
Before our results, the most compelling evidence for the $3D$ Ising transition being conformal was from numerical conformal bootstrap \cite{Rychkov:2009ij,ElShowk:2012ht,Kos:2016ysd,RMP_CB,Ising_CB}, which assumes conformal symmetry and found critical exponents close to the values obtained by various methods such as Monte Carlo simulation \cite{Hasenbusch2010,Landau2018} and measured by experiments \cite{Vicari2002}. 
In addition, there was an effort~\cite{Meneses2019viral} to justify the conformal invariance of the $3D$ Ising by showing that the virial current operator does not exist.~\footnote{We note that Ref.~\cite{Wschebor2016} claimed a proof of the conformal invariance of $3D$ Ising transition, but it is unclear if the proof is correct (see the comment in Appendix B in the first arXiv version of Ref.\cite{Meneses2019viral}).} 
Our obtained operator spectrum from the state-operator correspondence indeed convincingly shows that the $3D$ Ising transition does not have the virial current, which is a structural explanation of the $3D$ Ising being conformal~\cite{Nakayama2013ScaleVSConformal}.
A major surprise of our results is that an incredibly small system size ($8\sim 16$ total spins) is already enough to yield accurate conformal data of the $3D$ Ising CFT.
So we expect this approach to open a new avenue for studying higher dimensional phase transitions and CFTs. 
Firstly, there is a zoo of universalities that can be studied using our approach, which is amenable to various numerical techniques such as exact diagnolization (ED), density-matrix renormalization group (DMRG) and determinantal Monte Carlo.
This offers an opportunity to tackle many open questions regarding phase transitions, critical phases and CFTs.
Secondly, a number of new universal quantities can be computed once the $3D$ CFT is simulated on a sphere, such as operator product expansion coefficients, $F$ (of $F$-theorem)~\cite{Casini2011Towards,Jafferis2011,Myers2011Holographic,Casini2012Renormalization}, and the spherical binder ratio~\cite{Berkowitz2021Binder}, just to name a few.

The paper is organized as follows. 
In Sec. \ref{sec:review_CFT} we will review background knowledge including the radial quantization of CFTs and the state-operator correspondence. The spherical Landau level quantization and related fuzzy sphere are discussed in Sec. \ref{sec:review_QH}. Readers familiar with these topics can skip some of these subsections. 
In Sec. \ref{sec:model}, we formulate spherical Landau levels to regularize the $3D$ Ising transition on a fuzzy sphere. A global quantum phase diagram is presented. 
In Sec. \ref{sec:state-operator}, we present the low-lying energy spectra at the phase transition point, and analyze their one-to-one correspondence with the scaling operators  as predicted by the Ising CFT. This is the main result of this paper. 
At last, we present a discussion and outlook in Sec. \ref{sec:discussion}.

\begin{figure}[b]
\includegraphics[width=0.75\linewidth]{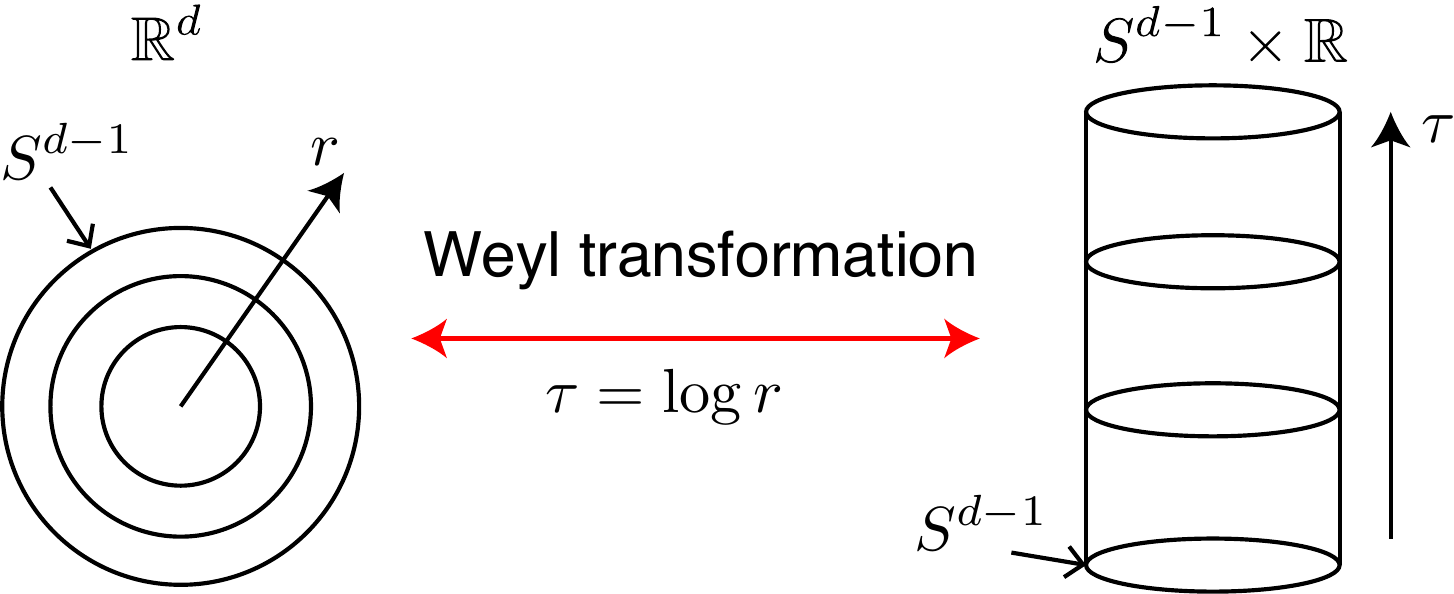}
\caption{Through a Weyl transformation, Euclidean flat space-time  $\mathbb R^d$ is mapped to the manifold of cylinder $S^{d-1}\times \mathbb R$. As a result, a CFT on $\mathbb R^d$ quantized on equal radius slices can be described equivalently in terms of a CFT on $S^{d-1}\times \mathbb R$ quantized on equal time slices. 
The states defined on the $S^{d-1}\times \mathbb R$  have well-defined quantum numbers of $SO(d)$ Lorentz rotation and dilatation, and thus they are in one-to-one correspondence with operators of the CFT, dubbed as \textit{state-operator correspondence}.}
\label{fig:radialQuan}
\end{figure}

\section{Review of background}

\subsection{ Radial quantization of CFTs: state-operator correspondence} \label{sec:review_CFT}

In this subsection we  review some basics of radial quantization, and for an elaborated discussion we refer the readers to CFT lecture notes such as those in~\cite{yellowbook,Rychkov2016lectures}. 

The conformal group in $d$ dimensions $SO(d+1,1)$ is generated by $d$-dimensional translations $P_\mu=i\partial_\mu$, $d$-dimensional Lorentz rotations $M_{\mu\nu}=i(x_\mu\partial_\nu - x_\nu \partial_\mu)$, dilatations $D=ix^\mu \partial_\mu$, and special conformal transformations $K_
\mu=i(2x_\mu(x^\nu \partial_\nu) - x^2 \partial_\mu)$.
From the operator point of view, a CFT can be thought of as a theory whose operators form an infinite-dimensional representation of the conformal group. 
Specifically, one can write CFT operators $\{ \hat O_\alpha \}$ as eigen-operators (i.e. irreducible representations) of the dilatation and Lorentz rotation $SO(d)$.
In particular, the eigenvalue $\Delta$ of dilatation is called scaling dimension of the operator, and it corresponds to the  exponent in the power law correlation function of the operator, e.g. $\langle O(x) O(0)\rangle \sim 1/|x|^{2\Delta}$.
One can further categorize  operators into primary operators and  descendant operators: 1) primary operators are operators that are annihilated by the special conformal transformation $K_\mu$; 2) descendant operators are not annihilated by $K_\mu$, and all of them can be obtained by applying translations $P_\mu$ (multiple times) to the primary operators.  
Therefore, one can organize CFT operators as primary operators and their descendants, and each primary and its descendants form a set of operators called a conformal multiplet.~\footnote{Here we are talking about primary operators under the global conformal symmetry $SO(d+1,1)$. For $2D$ CFTs one usually talks about primary operators under Virasoro symmetries, and the global conformal primaries are called quasi-primaries.}
A CFT  has an infinite number of primary operators, which makes it hard to tackle theoretically. 
A major task of solving a CFT is thus to obtain its low lying (if not full) spectrum of primary operators.

To facilitate later analysis of our numerical results,  we will elaborate a bit more about the operator contents of a $3D$ CFT. 
In $3D$ the Lorentz rotation group is the familiar $SO(3)$ group, all the irreducible representations of which are rank-$\ell$ symmetric traceless representations, i.e., spin-$\ell$ representations.
So all (primary and descendant) operators have two quantum numbers $(\Delta, \ell)$. 
A primary operator $O$ with quantum number $\ell=0$ is called a scalar operator, and any of its descendants can be written as 
\begin{equation}\label{eq:scalardesc}
\partial_{\nu_1} \cdots \partial_{\nu_j} \square^n O, \quad n, j\ge 0,
\end{equation}
with quantum number $(\Delta+2n+j, j)$. 
We note $\square = \partial^2$.
Here and hereafter all the free indices shall be symmetrized with the trace subtracted. 
The descendants of a spin-$\ell$ primary operator $O_{\mu_1 \cdots \mu_\ell}$ are a bit more complicated as there are two different types.
The first type can be written as,
\begin{equation}  \label{eq:spindesc1}
\partial_{\nu_1} \cdots \partial_{\nu_j} \partial_{\mu_1} \cdots \partial_{\mu_i} \square^n O_{\mu_1 \cdots \mu_\ell}, \quad
\end{equation}
with quantum number $(\Delta+2n+j+i, \ell+j-i)$ for $\ell \ge i\ge 0, \,\, n,j\ge 0$.
Here and hereafter the repeated indices shall be contracted.
The other type will involve the $\varepsilon$ tensor of $SO(3)$, and can be written as, 
\begin{equation} \label{eq:spindesc2}
\varepsilon_{\mu_l \rho \tau} \partial_\rho \partial_{\nu_1} \cdots \partial_{\nu_j} \partial_{\mu_1} \cdots \partial_{\mu_i} \square^n O_{\mu_1 \cdots \mu_\ell}, \quad 
\end{equation}
with quantum number $(\Delta+2n+j+i+1, \ell+j-i)$ for $\ell-1\ge i\ge 0, \,\, n,j\ge 0$.
We note that the $\varepsilon$ tensor alters spacetime parity symmetry of $O_{\mu_1 \cdots \mu_\ell}$.

We also remark that conserved operators (i.e. global symmetry current $J_\mu$ and energy momentum tensor $T_{\mu\nu}$)  should be treated a bit differently,  because they satisfy the conservation equations $\partial_{\mu} J_\mu=0$ and $\partial_{\mu} T_{\mu\nu}=0$. 
Therefore, their descendants in Eq.~\eqref{eq:spindesc1} and \eqref{eq:spindesc2}  should have $i=0$.~\footnote{The conformal multiplet of a conserved operator is called a short multiplet.}

Now we turn to the state perspective of CFTs. 
To define states of a CFT, we first need to quantize it, or in other words find a Hilbert space construction of it.
A quantum phase transition, namely a quantum Hamiltonian realization of a $d$-dimensional CFT in $d-1$ space dimensions, can be viewed as a way to quantize the CFT. 
The states of the CFT are nothing but the quantum Hamiltonian's eigenstates.
Formally, the quantization of CFTs can be more general than quantum phase transitions.
Specifically, one can foliate $d$-dimensional spacetime into $d-1$-dimensional surfaces, and each leaf of the foliation is endowed with its own Hilbert space. 
One convenient quantization is radial quantization, which has the $d$-dimensional Euclidean space $\mathbb{R}^d$ foliated to $S^{d-1}\times \mathbb{R}$, as shown in the left hand side of Fig.~\ref{fig:radialQuan}. 
In the radial quantization, the $SO(d)$ Lorentz rotation acts on the $S^{d-1}$ sphere, while the dilatation acts as the scaling of sphere radius.
Therefore, the states defined on the foliation $S^{d-1}$  have well-defined quantum numbers of $SO(d)$ rotation and dilatation, and they are indeed in one-to-one correspondence with operators of the CFT, dubbed as \textit{state-operator correspondence}. 

For a quantum Hamiltonian realization, the radial quantization described above is not natural, and instead one may want a quantization scheme that has an identical Hilbert space on each leaf of foliation. 
A quantum Hamiltonian is usually defined on the $M^{d-1}\times \mathbb{R}$ manifold: $\mathbb{R}$ is the time direction, while $M^{d-1}$ is a $d-1$-dimensional space manifold (e.g. sphere, torus, etc.), the leaf of foliation,  on which the Hilbert space (and the quantum state) lives. 
In order to discuss state-operator correspondence in such a quantization scheme, one needs  to map $\mathbb{R}^d$ to the cylinder $S^{d-1}\times \mathbb{R}$ using a Weyl transformation \cite{Cardy1984,CARDY1985}, as shown in Fig.~\ref{fig:radialQuan}. 
Under the Weyl transformation the dilatation $r\rightarrow e^\lambda r$ of $\mathbb{R}^d$  becomes the translation along the time direction $\tau \rightarrow \tau+\lambda$ of  $S^{d-1}\times \mathbb{R}$. 
If the theory has conformal symmetry, we can simply relate correlators and states on $\mathbb{R}^d$ to those on $S^{d-1}\times \mathbb{R}$. 
Moreover, we still have the state-operator correspondence on the cylinder $S^{d-1}\times \mathbb{R}$. 
In particular, the state-operator correspondence on the cylinder has a nice physical interpretation, namely the eigenstates $|\psi_n\rangle$ of the CFT quantum Hamiltonian on $S^{d-1}$ are in one-to-one correspondence with the CFT operators, and the energy gaps $\delta E_n$ of these states are proportional to the scaling dimensions $\Delta_n$ of CFT operators \cite{Cardy1984,CARDY1985},  
\begin{equation}
\delta E_n = E_n - E_0 = \frac{v}{R} \Delta_n,
\end{equation}
where $R$ is the radius of sphere $S^{d-1}$ and $v$ is the velocity of light that is model dependent. 
Also the $SO(d)$ rotation symmetry of $S^{d-1}$ is identified with the $SO(d)$ Lorentz rotation of the conformal group, so the $SO(d)$ quantum numbers of $|\psi_n\rangle$  are identical to those of CFT operators.

We emphasize that in contrast to radial quantization on $\mathbb{R}^d$, conformal symmetry is indispensable for the state-operator correspondence of radial quantization on the cylinder $S^{d-1}\times \mathbb{R}$.
Therefore, observing the state-operator correspondence on the cylinder $S^{d-1}\times \mathbb{R}$ will be direct evidence for the conformal symmetry of the theory or phase transition.
For $d=2$, the cylinder $S^1\times \mathbb{R}$ corresponds to nothing but a quantum Hamiltonian defined on a periodic chain, and there are very nice results studying  the resulting state-operator correspondence ~\cite{Blote1986Conformal,affleck1988universal,Milsted2017,Zou2018}. 
In higher dimensions, one needs to study a quantum Hamiltonian defined on $S^{d-1}$, however, it is highly nontrivial for a discrete lattice model as $S^{d-1\ge 2}$ has a curvature.

\subsection{Spherical Landau levels, fuzzy two-sphere and lowest Landau level projection}
\label{sec:review_QH}

As originally shown by Landau, electrons moving in $2D$ space under a magnetic field will form completely flat bands called Landau levels, 
which is the key to the quantum Hall effect. 
Landau level quantization can be considered on any orientable manifold, and Haldane~\cite{Sphere_LL_Haldane} first introduced Landau levels on spherical geometry to study the fractional quantum Hall physics. 

For electrons moving on the surface of a radius-$r$ sphere with a $4\pi s$ monopole ($2s\in \mathbb{Z}$) placed at the origin (Fig. \ref{fig:phase_diagram}),  the Hamiltonian is
\begin{equation} \label{eq:sphereQM}
H_0 = \frac{1}{2M_e r^2} \Lambda_\mu^2,
\end{equation}
where $M_e$ is the electron's mass and $\Lambda_\mu=\partial_\mu + i A_\mu$ is the covariant angular momentum, $A_\mu$ is the gauge field of the monopole. 
As usual we take $\hbar=e=c=1$.
The eigenstates will be quantized into spherical Landau levels, whose  energies are $E_n= [n(n+1)+(2n+1)s]/(2M_e r^2)$, with $n=0, 1, 2, \cdots$  the Landau level index.
The $(n+1)_\textrm{th}$ Landau level is $(2s+2n+1)$-fold degenerate, and
the single particle states in each Landau level are called Landau orbitals.
Assuming all interactions are much smaller than the energy gap between Landau levels, we can just consider the lowest Landau level (LLL) $n=0$, which is $2s+1$-fold degenerate.
The wave-functions  for each Landau orbital on LLL  are called monopole harmonics~\cite{WuYangmonopole}
\begin{equation} \label{eq:orbitals}
\Phi_{m} (\theta, \varphi)=N_m e^{im\varphi} \cos^{s+m}\left(\frac{\theta}{2}\right)\sin^{s-m}\left(\frac{\theta}{2}\right),
\end{equation}
with $m=-s, -s+1, \cdots, s$ and $N_m = \sqrt{\frac{(2s+1)!}{4\pi(s+m)!(s-m)!}}$.
Here $(\theta, \varphi)$ is the spherical coordinate. 

These LLL Landau orbitals indeed form a $SO(3)$ spin-$s$ irreducible representation. 
This can be understood by constructing the $SO(3)$ angular momentum operator \cite{Sphere_LL_Greiter},
\begin{equation}
L_\mu = \Lambda_\mu + s \frac{x_\mu}{r},
\end{equation}
which satisfies the $SO(3)$ algebra $[L_\mu, L_\nu] = i \varepsilon_{\mu\nu\rho} L_\rho$.
Projecting the system into the LLL, the kinetic energy of the covariant angular momentum will be quenched, so effectively we have $L_\mu \sim s \tilde{x}_\mu/r$. ($\tilde{x}_\mu$ denotes the coordinates in the projected LLL.)
As a result, the coordinates $\tilde x_\mu$ of electrons will not  actually commute, instead we have 
\begin{equation}
[\tilde x_\mu, \tilde x_\nu] =  i \frac{r}{s}   \epsilon_{\mu\nu\rho}   \tilde x_\rho.
\end{equation}
This defines a fuzzy two-sphere~\cite{madore1992fuzzy}. 
Moreover, Landau orbitals~\eqref{eq:orbitals}  are in one-to-one correspondence with states on the fuzzy two-sphere. 
Formally, a system defined on the LLL can be equivalently viewed as a system defined on a fuzzy two-sphere.
We will not delve into details along that direction, and refer the reader to \cite{Hasebe2010fuzzy} for more discussions. 

As is usually done in the literature, we will consider the limit where the interaction strength is much smaller than the Landau level gap, so we can project the system into the LLL. 
Technically, this can be done by rewriting the annihilation operator $\psi(\theta, \varphi)$ on the LLL as
\begin{equation}
\hat \psi(\theta, \varphi) = \frac{1}{\sqrt{2s+1}}  \sum_{m=-s}^s \Phi^*_m \hat c_m.
\end{equation}
$\hat c_m$ stands for the annihilation operator of Landau orbital $m$, and is independent of coordinates $(\theta, \varphi)$.
The density operator $n(\theta, \varphi) = \psi^\dag \psi$ can be written as,
\begin{equation}
n(\theta, \varphi) = \frac{1}{2s+1} \sum_{m_1, m_2} \Phi_{m_1} \Phi^*_{m_2} c^\dag_{m_1} c_{m_2}.
\end{equation}
Any interaction can be straightforwardly (though perhaps tediously) written in the second quantized form using Landau orbital operators $c^\dag_m, c_m$.
For example, the density-density interaction $H_I= \int d^2 \vec r_a d^2 \vec r_b \; U(\vec r_a -\vec r_b) n (\vec r_a)  n (\vec r_b)$ can be written as,
\begin{align}\label{eq:LLL_dd}
H_I &=  (2s+1)^2 \int d\Omega_a d\Omega_b \, U(\theta_a, \varphi_a; \theta_b, \varphi_b) n(\theta_a, \varphi_a)  n(\theta_b, \varphi_b)  \nonumber \\ 
& = \sum_{m_1, m_2, m_3, m_4} V_{m_1, m_2, m_3, m_4} \, c^\dag_{m_1} c^\dag_{m_2} c_{m_3}  c_{m_4},
\end{align}
where $V_{m_1, m_2, m_3, m_4}$ can be further expanded using the so-called Haldane pseudopotential  $V_l$ \cite{Sphere_LL_Haldane}, corresponding to the two-fermion scattering in the spin-$2s-l$ channel (see Appendix Sec. \ref{app:pseudopotential}). 

In summary, the model we are working with is a fermonic Hamiltonian enclosing $2s+1$-Landau orbitals with long-range $SO(3)$ invariant interactions. 
Interestingly, all the orbitals form  an $SO(3)$ spin-$s$ irrep.
Furthermore, the length scale of the system is $\sqrt{2s+1}$ instead of $2s+1$ since the spatial dimension is $d=2$, and the thermodynamic limit corresponds to taking $s$ to infinity.

\section{Model on a fuzzy-two sphere}
\label{sec:model}

\subsection{Hamiltonian}

\begin{figure}
	\includegraphics[width=0.95\linewidth]{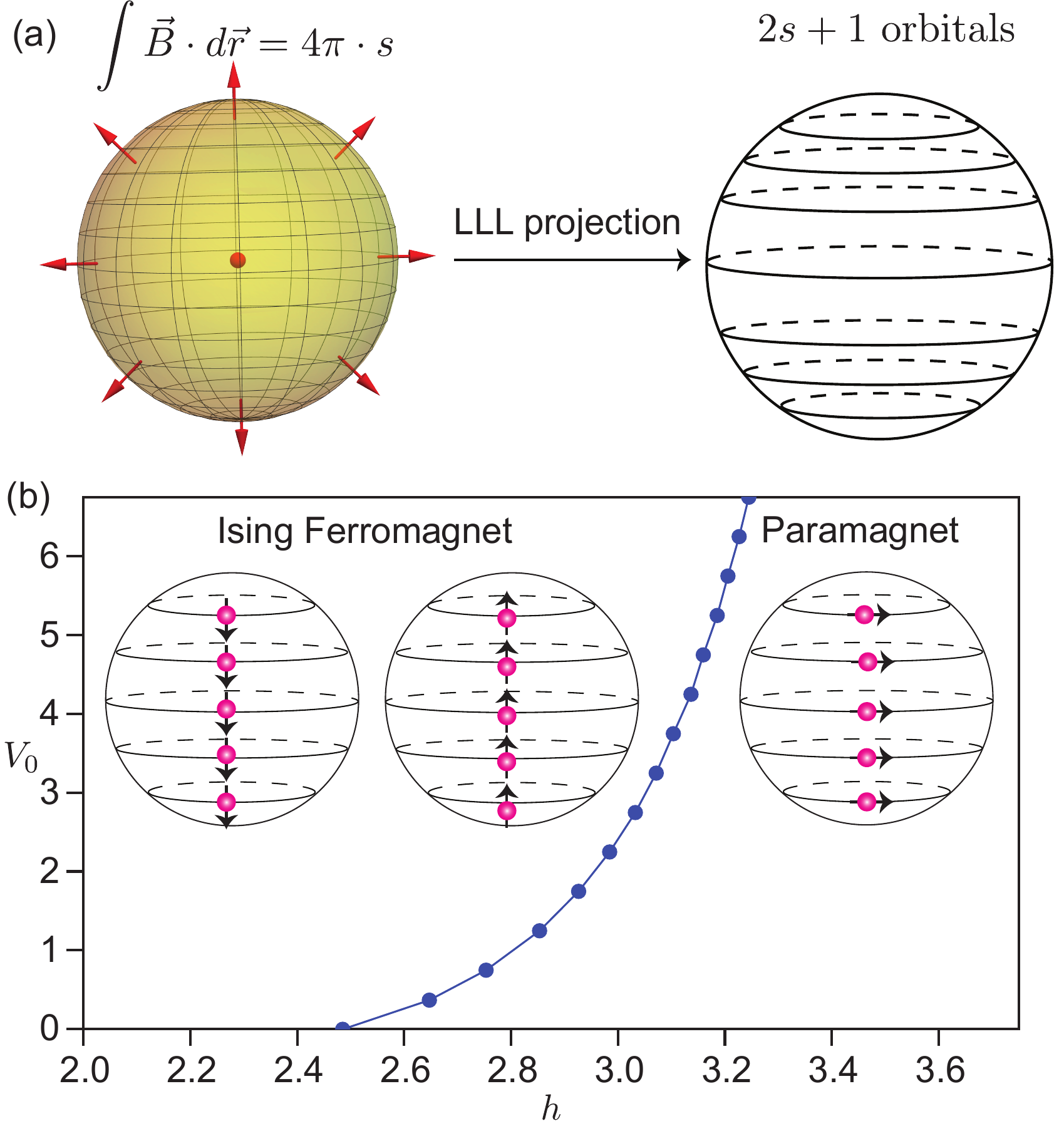}
	\caption{(a) Schematic plot of electrons moving on a sphere in the presence of $4\pi \cdot s$ monopole. The LLL has $2s+1$ degenerate orbitals, which form an $SO(3)$ spin-$s$ irreducible representation. A system projected into the LLL can be equivalently viewed as a fuzzy sphere.  (b) Phase diagram of the proposed model  consisting of a continuous phase transition from a quantum Hall Ising ferromagnet to a disordered paramagnet.}
	\label{fig:phase_diagram}
\end{figure}

\begin{widetext}

Here we explicitly define the model, which is spinful electrons in the LLL.~\footnote{The spin degree of freedom should be thought as a pseudospin as it does not couple to the Zeeman field of the magnetic monopole.}
In spatial space, the Hamiltonian takes the form
\begin{align}
H &=   \int (2s+1)^2 d\Omega_a d\Omega_b \,  U(\Omega_{ab}) \left[n^0(\theta_a, \varphi_a)n^0(\theta_b, \varphi_b)- n^z(\theta_a, \varphi_a)n^z(\theta_b, \varphi_b) \right] - h  \int (2s+1) d \Omega \, n^x(\theta,\varphi), 
\end{align}
where $n^\alpha (\theta,\varphi)$ is a local density operator given by
\begin{equation}
n^\alpha (\theta,\varphi) = ( \hat\psi^\dag_{\uparrow}(\theta,\varphi), \, \hat\psi^\dag_{\downarrow}(\theta,\varphi) ) \, \sigma^\alpha \left(\begin{matrix}
\hat\psi_{\uparrow}(\theta,\varphi) \\ 
\hat\psi_{\downarrow}(\theta,\varphi),
\end{matrix} \right),
\end{equation}
with $\sigma^{x,y,z}$ being Pauli matrices, $\sigma^0=I_{2\times 2}$, and $U(\Omega_{ab})$ the local density-density interactions (defined below). The first term behaves like an Ising ferromagnetic interaction, while the second term is the transverse field. 
By projecting the Hamiltonian into the LLL, we  obtain 
\begin{align} \label{eq:HamIsing}
\begin{split}
H & = H_{00}+H_{zz} + H_t, \\ 
H_{00} &= \frac{1}{2} \sum_{m_{1,2,3,4}=-s}^s    V_{m_1,m_2,m_3,m_4} \left(\mathbf{c}_{m_1}^\dag \mathbf{c}_{m_4} \right) 
\left( \mathbf{c}_{m_2}^\dag  \mathbf{c}_{m_3} \right) 
 \delta_{m_1+m_2,m_3+m_4} , \\ 
H_{zz} &= - \frac{1}{2}  \sum_{m_{1,2,3,4}=-s}^s   V_{m_1,m_2,m_3,m_4} \left(\mathbf{c}_{m_1}^\dag \sigma^z \mathbf{c}_{m_4} \right) \left( \mathbf{c}_{m_2}^\dag \sigma^z \mathbf{c}_{m_3} \right) \delta_{m_1+m_2,m_3+m_4}, \\ 
H_t & = -h \sum_{m=-s}^s \mathbf{c}_m^\dag \sigma^x \mathbf{c}_m,
\end{split}
\end{align}
where $\mathbf{c}^\dag_m = (c^\dag_{m\uparrow}, c^\dag_{m\downarrow})$ is the fermion creation operator on the $m_\textrm{th}$ Landau orbital.
The parameter $V_{m_1,m_2,m_3,m_4}$ is connected to the Haldane pseudopotential $V_l$ by
\begin{align}
V_{m_1, m_2, m_3, m_4} 
&= \sum_l V_l \, (4s-2l+1)\tj{s}{s}{2s-l}{m_1}{m_2}{-m_1-m_2} 
\tj{s}{s}{2s-l}{m_4}{m_3}{-m_3-m_4},
\end{align} 
where $\tj{j_1}{j_2}{j_3}{m_1}{m_2}{m_3}$ is the Wigner $3j$-symbol. 
In this paper we will only consider ultra-local  density-density interactions in real space, i.e. $U(\Omega_{ab}) = g_0 \frac{1}{R^2} \delta(\Omega_{ab}) +g_1 \frac{1}{R^4} \nabla^2\delta(\Omega_{ab})$, and the associated Haldane pseudopotentials involve $V_0,V_1$ (see Appendix Sec. \ref{app:pseudopotential}). Next we will set $V_1=1$ as energy unit and vary $V_0,h$ to study the phase diagram. 

We consider the half-filling case with the LLL filled by $N=2s+1$ electrons. 
When $h=0$ and $V_{0},V_1>0$, the ground state is an Ising ferromagnet that spontaneously breaks $\mathbb{Z}_2$ symmetry. 
In quantum Hall literature this phase is called quantum Hall ferromagnetism \cite{Sondhi1993,Girvin2000}. 
The two-fold degenerate ground states are $|\Psi_{\uparrow}\rangle = \prod_{m=-s}^s c^\dag_{m\uparrow} |0\rangle $ and $|\Psi_{\downarrow}\rangle = \prod_{m=-s}^s c^\dag_{m\downarrow} |0\rangle $. 
When $h \gg V_{0},V_1$, the ground state is a trivial paramagnet that preserves Ising symmetry, $|\Psi_{x}\rangle = \prod_{m=-s}^s (c^\dag_{m\uparrow}+c^\dag_{m\downarrow}) |0\rangle $. Therefore, we expect a $2+1D$ Ising transition as increasing $h$. 
The global phase diagram of the model is as shown in Fig.~\ref{fig:phase_diagram}(b).

\end{widetext}

\subsection{Symmetries and order parameter}

The Hamiltonian~\eqref{eq:HamIsing} has three symmetries,
\begin{enumerate}
    \item Ising $\mathbb{Z}_2$ symmetry: $\mathbf{c}_m \rightarrow \sigma^x \mathbf{c}_m$.
    \item $SO(3)$ sphere rotation symmetry: $\mathbf{c}_{m=-s,\cdots, s}$ form the spin-$s$ representation of $SO(3)$.
    \item Particle-hole symmetry: $\mathbf{c}_m \rightarrow i\sigma^y \mathbf{c}_m^*$, $i\rightarrow -i$.
\end{enumerate}
Electric charges of fermions are gapped in the entire phase diagram (see Appendix Sec. \ref{smsec:fermiongap}), while the Ising spins of fermions are the degrees of freedom that go through the phase transitions.
Therefore, all the gapless degrees of freedom at the phase transition are charge-neutral. 
In particular, the order parameter of the transition is a particle-hole excitation of fermions,
\begin{equation} \label{eq:orderp}
M = \sum_{m=-s}^s  \mathbf{c}^\dag_m \frac{\sigma^z}{2} \mathbf{c}_m.
\end{equation}
We emphasize an important point for the Landau level regularization of the Ising transition: the electrons are sitting on a fuzzy sphere due to the monopole, but the Ising spins are sitting on a normal sphere (for any finite $N=2s+1$) since they are charge neutral. 
This is the key difference between our Landau level regularization and the non-commutative field theory~\cite{noncommuQFT}, namely the latter always has quantum fields defined on a fuzzy manifold as long as the physical volume is finite. 

To further analyze the Ising transition in our system, we  will relate the UV symmetries of our Landau level model to the IR symmetries of the 3D Ising CFT. 
It is obvious we can identify the Ising $\mathbb{Z}_2$ and $SO(3)$ sphere rotation between UV and IR.
A slightly non-trivial symmetry is the particle-hole symmetry, which turns out to be the spacetime parity symmetry of 3D Ising CFT. 
To understand this relation, we can write an $SO(3)$ vector, 
\begin{equation}
n_{m=0,\pm1}^x = \sum_{m_1=-s}^s (-1)^{m_1} \tj{s}{s}{1}{m_1}{m-m_1}{-m} \mathbf{c}^\dag_{m_1} \sigma^x \mathbf{c}_{m_1-m},
\end{equation}
and find it transforms as
\begin{equation}
\left( \begin{matrix}
n^x_{m=1} \\ n^x_{m=0}  \\ n^x_{m=-1} 
\end{matrix} \right) \rightarrow  
\left(
\begin{matrix}
0 & 0 & -1 \\
0 & 1 & 0 \\
-1 & 0 & 0
\end{matrix} \right) \left( \begin{matrix}
n^x_{m=1} \\ n^x_{m=0}  \\ n^x_{m=-1} 
\end{matrix} \right),
\end{equation}
under particle-hole transformation.
The particle-hole acts as an improper $\mathbb{Z}_2$ of $O(3)$, so it can be identified as the spacetime parity of the 3D Ising CFT. 

\subsection{Finite size scaling}

\begin{figure}
    \centering
\includegraphics[width=0.95\linewidth]{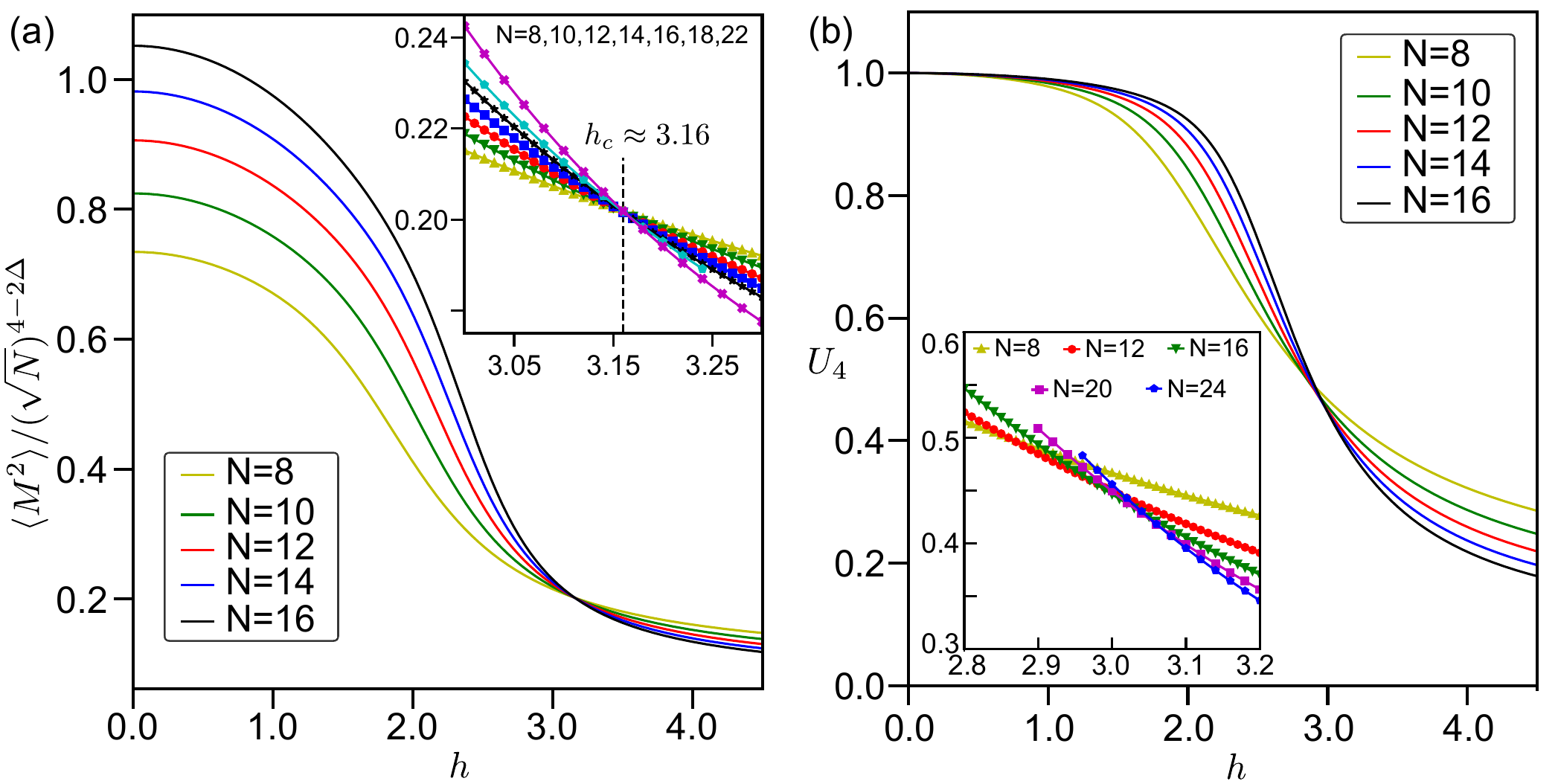}
    \caption{(a) Finite size scaling of order parameter $\langle M^2 \rangle/N^{2-\Delta}$. $\Delta=0.518148$ is the scaling dimension of the Ising order parameter field. $N=2s+1$ is the number of electrons (i.e. Ising spins), hence it should be identified as space volume and the length scale is $\sim\sqrt N$.
    The rescaled order parameter perfectly crosses at the same point $h_c\approx 3.16$. (b) Plot of the RG-invariant binder cumulant $U_4$. The binder cumulant does not stably cross at the same point due to the large finite size effect. We set $V_0=4.75$ here. 
    }
    \label{fig:scaling}
\end{figure}

The phase diagram in Fig.~\ref{fig:phase_diagram}(b) is obtained by the conventional finite size scaling of the $\mathbb{Z}_2$ order parameter $M$ in Eq.~\eqref{eq:orderp}.
We have simulated $N=2s+1=8,10,\cdots 24$ using ED for smaller sizes ($N\le 16$) and DMRG for larger sizes $N>16$ (the length scale in this $2+1$D system is $L_x=\sqrt{N}$). 
At the phase transition point, the $\mathbb{Z}_2$ order parameter should scale as $\langle M^2 \rangle \sim L_x^{4-2\Delta} = N^{2- \Delta}$ \cite{Hasenbusch2010},
where $\Delta\approx 0.5181489$ is the scaling dimension of Ising order parameter \cite{RMP_CB,Ising_CB}. 
Fig.~\ref{fig:scaling} (a) depicts $\langle M ^2 \rangle/N^{2-\Delta}$ with respect with the transverse field strength $h$ of different $N$ for $V_0 = 4.75$.
All the curves nicely cross at $h_c\approx 3.16$, which we identify as the transition point. 
Similarly for other $V_0$ we have identified the critical $h_c$ and obtained the phase diagram as shown in Fig.~\ref{fig:phase_diagram}(b).

We have also computed the binder cumulant 
\begin{equation}
U_4 = \frac{3}{2} \left( 1 - \frac{1}{3}\frac{\langle M^4 \rangle }{\langle M^2 \rangle^2 }\right).
\end{equation}
$U_4$ is a RG-invariant quantity, and $U_4=1,0$ at the thermodynamic limit corresponds to the ordered phase and disordered phase, respectively. 
At the phase transition $U_4$ will be a universal quantity related to the four point correlator of the order parameter field $\sigma$ of CFT~\cite{Berkowitz2021Binder}. 
Fig.~\ref{fig:scaling} (b) shows $U_4$ with respect  to the transverse field strength $h$  for different $N$ for $V_0 = 4.75$. Clearly, at small $h$ the model is in the Ising ferromagnetic phase, while at large $h$ the model is in the disordered phase. 
To estimate the value of binder ratio at the critical point $U^c_4$, we perform a detailed crossing-point analysis (Appendix Sec. \ref{smsec:M2}). With the data on hand, the best estimate we can give is $0.28\le U^c_4 \le 0.40$. It will be interesting to evaluate $U_4$ from conformal bootstrap and compare with our estimate. 
~\footnote{For models on the non-conformal manifold such as $T^2\times \mathbb{R}$ or $T^3$, which Monte Carlo usually simulates, $U_4$ cannot be computed using the $R^3$ four-point correlator from conformal bootstrap.}

In practice, for small $N$ (as we simulated numerically), finite-size effects are inevitable. One common source is from the couplings of irrelevant operators, which are typically present in microscopic models. Tuning along the critical line in the 2-dimensional parameter space $(V_0,h)$ shown in Fig. \ref{fig:phase_diagram}(b) generically modifies the coupling strength of irrelevant operators and therefore the magnitude of finite-size effects (while the relevant operators flow to the same fixed point).
In the following section, we will present the data of the state-operator correspondence at a particular point $V_0=4.75, h_c=3.16$, where we find the finite size effects are smallest (Appendix Sec. \ref{smsec:conformaldata}).

\section{State-operator correspondence}
\label{sec:state-operator}

We now turn to the central results of our paper: the state-operator correspondence of the $3D$ Ising transition.
As explained in Sec.~\ref{sec:review_CFT}, on $S^2\times \mathbb{R}$ the eigenstates of the quantum Hamiltonian are in one-to-one correspondence with the scaling operators of its corresponding CFT. 
In particular, the energy gaps of each state will be proportional to the scaling dimensions of the scaling operators \cite{CARDY1985}. 
Therefore, we explore energy spectra at the critical point by utilizing exact diagonalization and compare it with CFT predictions. 

To match the Ising transition's energy spectra with the $3D$ Ising CFT's operator spectrum, we first need to rescale the energy spectrum with a non-universal (i.e. model- and size-dependent) numerical factor. 
The natural calibrator is the energy momentum tensor $T_{\mu_1\mu_2}$, a conserved operator that any local CFT possesses. 
For any $3D$ CFT, $T_{\mu_1\mu_2}$ will be a global symmetry singlet, Lorentz spin $\ell=2$ operator with scaling dimension $\Delta_T=3$.
Our model has exact $SO(3)$ Lorentz rotation, Ising $\mathbb{Z}_2$, and spacetime parity symmetries, so every eigenstate has well-defined quantum numbers $(\mathbb{Z}_2, P, \ell)$ of these three symmetries.  
The energy-momentum tensor will be the lowest state in the $(\mathbb{Z}_2=1, P=1, \ell=2)$ sector. 
We rescale the full spectrum by setting the energy momentum tensor to exactly $\Delta_T=3$, and then examine if the low-lying states form representations of $3D$ conformal symmetry  up to a finite size correction.

\begin{table*}
\setlength{\tabcolsep}{0.2cm}
\renewcommand{\arraystretch}{1.4}
    \centering
    \caption{Low-lying primary operators identified via state-operator correspondence on a fuzzy sphere with $N=16$ electrons. The operators in the first and second row are $\mathbb{Z}_2$ odd and even operators, respectively. We highlight that two new parity-odd primary operators $\sigma^{P-}$ and $\epsilon^{P-}$ are found. The conformal bootstrap data is from Ref. \cite{Ising_CB}.} \label{tab:primary}
\begin{tabular}{cccccccc|c} \hline\hline
& $\sigma$ & $\sigma'$  & $\sigma_{\mu_1 \mu_2}$ & $\sigma'_{\mu_1 \mu_2}$ & $\sigma_{\mu_1 \mu_2 \mu_3}$ & $\sigma_{\mu_1 \mu_2 \mu_3 \mu_4}$ & & $\sigma^{P-}$\\
Bootstrap  & 0.518 & 5.291 & 4.180  & 6.987  & 4.638 & 6.113 && NA \\
Fuzzy sphere & 0.524 & 5.303 & 4.214 & 7.048 & 4.609  & 6.069 && 11.191 \\
\hline 
& $\epsilon$ & $\epsilon'$  & $\epsilon''$  & $T_{\mu\nu}$ & $T'_{\mu\nu}$ & $\epsilon_{\mu_1\mu_2\mu_3\mu_4}$ & $\epsilon'_{\mu_1\mu_2\mu_3\mu_4}$ & $\epsilon^{P-}$ \\
Bootstrap  & 1.413 & 3.830  & 6.896 & 3 & 5.509 & 5.023 &  6.421 & NA  \\
Fuzzy sphere & 1.414 & 3.838 & 6.908  & 3 & 5.583 & 5.103 &  6.347 & 10.014 \\ 
 \hline\hline

\end{tabular} 
\end{table*}

\begin{figure*}
    \centering
\includegraphics[width=0.95\textwidth]{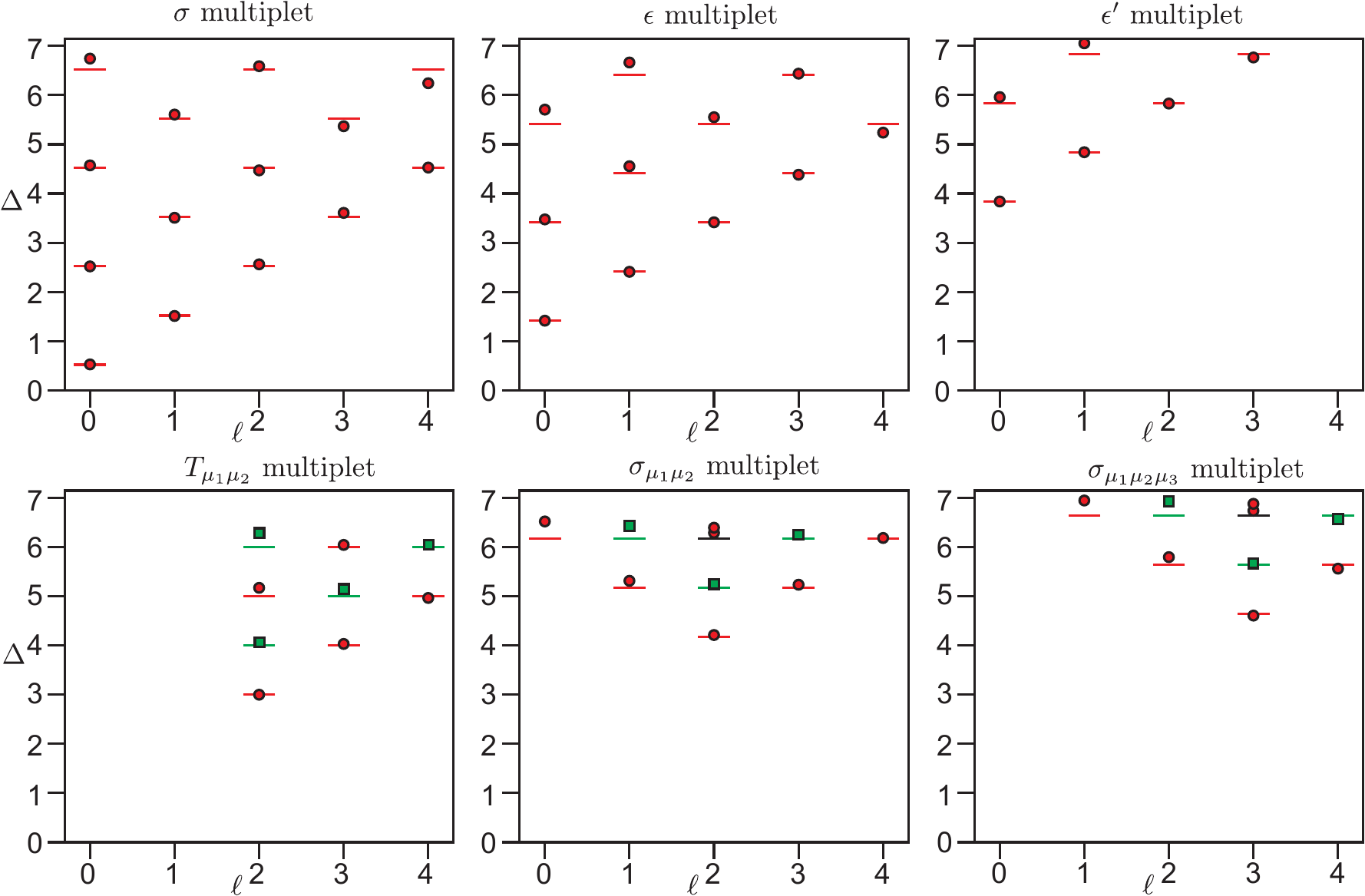}
    \caption{Conformal multiplet of several low lying primary operators: scaling dimension $\Delta$ versus Lorentz spin $\ell$. We plot conformal bootstrap data with lines: lines in red are parity even, non-degenerate operators;  lines in green are parity odd, non-degenerate operators;  lines in black are parity even, two-fold degenerate operators. Symbols are our numerical data of parity even (red circle) and odd (green square) operators. 
    The discrepancy is typically more significant for the larger $\Delta$.
    }
    \label{fig:multiplet}
\end{figure*}

We analyze the low-lying spectra according to the following steps,
\begin{enumerate}
    \item For each $\mathbb Z_2=\pm 1$ sector, we find the lowest-lying energy state (regardless of $\ell$ and $P$), and identify it as a primary state. 
    \item Based on the representation theory of the 3D conformal group as summarized in Eq.~\eqref{eq:scalardesc}, \eqref{eq:spindesc1}, \eqref{eq:spindesc2}, we enumerate the descendant states of the identified primary state and examine if all of  descendant states (up to $\Delta=7$) exist in our energy spectrum. 
    \item We remove the identified conformal multiplet (i.e. primary and its descendants) from the energy spectrum, and for the remaining states we repeat the step 1,2.
\end{enumerate}
Remarkably, we found that the lowest-lying 70 eigenstates~\footnote{We have targeted the lowest $100$ eigenstates using ED without explicitly imposing the value of $\ell$, and we only looked at states with $\ell\le 4$ which roughly contains $70$ states.} form representations of the 3D conformal symmetry up to a small finite size correction, with no extra or missing state.
This is a direct and unambiguous demonstration of the emergent conformal symmetry of the $3D$ Ising transition.

After verifying the emergent conformal symmetry, we further compare our  scaling dimensions of the identified primary operators with the numerical conformal bootstrap data~\cite{Ising_CB,RMP_CB}, and we find a good agreement for all of them.
Table~\ref{tab:primary} lists all the primary operators we have identified with $N=16$ ED data.
We have found 12 parity-even primary operators besides the energy-momentum tensor, and all of them  have less than a $1.6\%$ discrepancy from the bootstrap data \cite{Ising_CB,RMP_CB}. 
In Appendix Sec. \ref{smsec:conformaldata} we list concrete values of each conformal multiplet, as one can see the numerical accuracy is unexpectedly high, particularly given that it is from a small system size ($N=16$ total spins): around 10 operators have relative numerical error around $3\%\sim 5.5\%$, and the rest of them have relative numerical error smaller than $3\%$.
Fig.~\ref{fig:multiplet} plots conformal multiplets of a few representative primary operators, which clearly illustrate the emergent conformal symmetry and agree well with numerical conformal bootstrap results.

A few remarks are in order. 
1) We verify the emergent conformal symmetry of the $3D$ Ising transition by showing that the low-lying spectra of our model form representations of $3D$ conformal symmetry. This procedure does not rely on any input of previous knowledge such as numerical bootstrap data.
2) A spinning ($\ell>0$) parity-even (parity-odd)  primary operator can have parity-odd (parity-even) descendant opertors as written in Eq.~\eqref{eq:spindesc2}. 
This nontrivial structure from the CFT's algebra matches our ED spectrum.~\footnote{To recall, the UV particle-hole symmetry becomes the spacetime parity symmetry of the IR CFT.}
3) The energy momentum tensor $T_{\mu_1\mu_2}$ is a conserved operator, so it does not have any $\ell<2$ descendant.
This structure is clearly shown in our data.
4) All the parity-even primary operators that we found have been reported in the bootstrap study of mixed correlators $\langle \sigma \sigma \sigma \sigma \rangle$, $\langle \epsilon \epsilon \epsilon \epsilon \rangle$,  $\langle \sigma \sigma \epsilon \epsilon \rangle$.
The mixed-correlator bootstrap study is only capable of detecting operators in the $\sigma \times \sigma$, $\epsilon\times \epsilon$ and $\sigma\times \epsilon$ OPE, so it will miss $(\mathbb{Z}_2=1, P=1, \textrm{odd} \, \ell)$ primary operators (in addition to $P=-1$ primaries). 
Our approach should be able to detect operators in these quantum number sectors, including the candidate of virial current~\cite{polchinski1988scale}~\footnote{Strictly speaking, virial current refers to an operator with scaling dimension $\Delta=2$. If such an operator exists, one may have a theory that is scale-invariant but not conformal-invariant.}, namely the lowest primary in the $(\mathbb{Z}_2=1, P=1, \ell=1)$ sector.
We have not observed any primary operators in the  $(\mathbb{Z}_2=1, P=1, \textrm{odd} \, \ell)$ sector  below $\Delta=7$, and so this gives a lower bound for the virial current candidate, which is higher than the previous estimate~\cite{Meneses2019viral}.
5) We have identified two previously unknown (parity-odd) primary operators in the $(\mathbb{Z}_2=1, P=-1, \ell=0)$ and $(\mathbb{Z}_2=-1, P=-1, \ell=0)$ sectors with $\Delta\approx 10.01$ and $\Delta\approx 11.19$, respectively.
To access $P=-1$ primary operators in the bootstrap calculation, one has to bootstrap correlation functions of the spinning operator: for example, the energy momentum tensor. 
Such study has only been initiated in Ref.~\cite{EMT_boot} but no $P=-1$ primary has been identified by conformal bootstrap or any other methods so far.
6) In all previous lattice model studies, only several primary fields ($\sigma$, $\epsilon$ and $\epsilon'$) were found, and their scaling dimensions are related to the critical exponents $\eta$, $\nu$ and $\omega$~\cite{Hasenbusch2010,Landau2018}.

\section{Summary and discussion}
\label{sec:discussion}

We  have designed an innovative scheme to numerically study the $3D$ Ising transition on the space-time geometry $S^2\times \mathbb{R}$, and in our calculation we  have found almost perfect state-operator correspondence of the $3D$ CFT, supporting the conjecture that the $3D$ Ising transition has emergent conformal symmetry.
In detail, we  considered the $3D$ Ising transition realized in a fermionic model defined on a fuzzy sphere,  which we  achieved by projecting spinful electrons into the lowest spherical Landau level where the spin degrees of freedom go through an order-disorder transition. 
We are able to identify 13 parity-even primary operators and 2 parity-odd primary operators, and around 60 descendant operators, in agreement with the predictions of underlying CFT within a high accuracy. 

Our results have now offered a novel solution to the long-standing quest of simulating $3D$ CFTs on the sphere (more generally on the curved space), and even more remarkably, the finite size effect of our model is much smaller than the conventional approach (i.e. 3D classical Ising model) used to study $3D$ CFTs. 
Therefore, our results open a new avenue for studying $3D$ CFTs in a microscopic way.
Thanks to the state-operator correspondence on $S^2\times \mathbb R$, many universal quantities such as operator product expansion coefficients, four-point correlators, and thermal correlators of CFTs are ready to compute directly. 
These will lead to many insights of CFTs that are important for several purposes. 
For example, the thermal correlators will not only be useful for predicting experiments of $2+1$ dimensional quantum phase transitions at the finite temperature, but also help to understand the properties of quantum black holes using the AdS/CFT duality.~\footnote{A CFT at finite temperature on boundary is dual to a black hole of quantum gravity in the bulk.} 
Another interesting quantity is the RG monotonic quantity $F$ of the $F$-theorem~\cite{Casini2011Towards,Jafferis2011,Myers2011Holographic,Casini2012Renormalization}, which can be extracted from the quantum entanglement~\cite{Klebanov2012Etanglement,Casini2015Mutual}.

We also expect our approach can be used to tackle many open problems of the $3D$ CFTs. 
Specifically, our approach can be applied to many  universalities such as $O(N)$ Wilson-Fisher transitions (i.e. XY universality, etc.) and critical gauge theories (e.g. see Ref.~\cite{Wang2021SO5WZW}).~\footnote{A guiding principle for the model design will be kinematical properties such as symmetries and anomalies.
Here we start with spinful fermions, whose maximal spin symmetry is $SU(2)$. We further add interactions that break $SU(2)$ down to the anomaly-free Ising $\mathbb Z_2$ symmetry. Similarly, for other transitions one can start with $N-$component fermions, and consider various interactions that break the maximal $SU(N)$ symmetry to the desired symmetries.} 
With a straightforward examination of the emergent conformal symmetry and a precise determination of the scaling dimensions of various primary operators, one may eventually solve the question of the conformal window of $3D$ critical gauge theories, a problem puzzled the high energy physics and condensed matter community for decades.

In our paper the fuzzy sphere regularization is formulated using the language of lowest Landau level projection.
It will be interesting to translate our formulation into the formal language of non-commutative geometry. 
Such perspective of the fuzzy sphere regularization
may help to develop a systematic framework that is applicable to any CFT and QFT on various manifolds in arbitrary space-time dimensions. 
For example, an ambitious question is, can one directly regularize the continuum QFTs on the fuzzy sphere without encountering the infamous UV infiniteness of QFTs?
Indeed the similar idea was pursued decades ago in the context of non-commutative field theory~\cite{noncommuQFT}, but was unsuccessful due to the phenomenon called UV-IR mixing.
Our regularization scheme offers a new angle to this question, namely one can introduce auxiliary fields (i.e. electrons in our model) that are living on the fuzzy sphere, and the true low energy quantum fields of the theory (i.e. Ising spins in our model) are living on the normal sphere. 
We believe this way of thinking may lead to many fruitful results of CFTs and QFTs, and may reveal a new connection between physics and mathematics.

\section{Acknowlegement}
We thank Chong Wang, Duncan Haldane, Sung-Sik Lee, Rob Myers, Junchen Rong, Yijian Zou for useful discussions. We thank Liangdong Hu for simulation discussion and collaboration on a related project. This work was supported by National Science Foundation of China under No. 92165102, 11974288 and National R\&D program under No. 2022YFA1402204 (W.Z., C.H.). Research at Perimeter Institute is supported in part by the Government of Canada through the Department of Innovation, Science and Industry Canada and by the Province of Ontario through the Ministry of Colleges and Universities.
J.H. was supported by the European Research Council (ERC) under grant HQMAT (Grant Agreement No. 817799), the Israel-US Binational Science Foundation (BSF), and by a Research grant from Irving and Cherna Moskowitz.
Y.C.H. and E.H.  thank Centre de recherches mathématiques (CRM) and the organizers of the workshop ``Conformal field theory and quantum many-body physics" for hospitality where some ideas of this work  were initiated, and for the Galileo Galilei Institute and the organizers of the GGI workshop ``Bootstrapping Nature: Non-perturbative Approaches to Critical Phenomena” for hospitality during the completion of this work.

% references
\bibliography{ref}

%merlin.mbs apsrev4-1.bst 2010-07-25 4.21a (PWD, AO, DPC) hacked
%Control: key (0)
%Control: author (0) dotless jnrlst
%Control: editor formatted (1) identically to author
%Control: production of article title (0) allowed
%Control: page (1) range
%Control: year (0) verbatim
%Control: production of eprint (0) enabled
\begin{thebibliography}{49}%
\makeatletter
\providecommand \@ifxundefined [1]{%
 \@ifx{#1\undefined}
}%
\providecommand \@ifnum [1]{%
 \ifnum #1\expandafter \@firstoftwo
 \else \expandafter \@secondoftwo
 \fi
}%
\providecommand \@ifx [1]{%
 \ifx #1\expandafter \@firstoftwo
 \else \expandafter \@secondoftwo
 \fi
}%
\providecommand \natexlab [1]{#1}%
\providecommand \enquote  [1]{``#1''}%
\providecommand \bibnamefont  [1]{#1}%
\providecommand \bibfnamefont [1]{#1}%
\providecommand \citenamefont [1]{#1}%
\providecommand \href@noop [0]{\@secondoftwo}%
\providecommand \href [0]{\begingroup \@sanitize@url \@href}%
\providecommand \@href[1]{\@@startlink{#1}\@@href}%
\providecommand \@@href[1]{\endgroup#1\@@endlink}%
\providecommand \@sanitize@url [0]{\catcode `\\12\catcode `\$12\catcode
  `\&12\catcode `\#12\catcode `\^12\catcode `\_12\catcode `\%12\relax}%
\providecommand \@@startlink[1]{}%
\providecommand \@@endlink[0]{}%
\providecommand \url  [0]{\begingroup\@sanitize@url \@url }%
\providecommand \@url [1]{\endgroup\@href {#1}{\urlprefix }}%
\providecommand \urlprefix  [0]{URL }%
\providecommand \Eprint [0]{\href }%
\providecommand \doibase [0]{http://dx.doi.org/}%
\providecommand \selectlanguage [0]{\@gobble}%
\providecommand \bibinfo  [0]{\@secondoftwo}%
\providecommand \bibfield  [0]{\@secondoftwo}%
\providecommand \translation [1]{[#1]}%
\providecommand \BibitemOpen [0]{}%
\providecommand \bibitemStop [0]{}%
\providecommand \bibitemNoStop [0]{.\EOS\space}%
\providecommand \EOS [0]{\spacefactor3000\relax}%
\providecommand \BibitemShut  [1]{\csname bibitem#1\endcsname}%
\let\auto@bib@innerbib\@empty
%</preamble>
\bibitem [{\citenamefont {Polyakov}(1970)}]{polyakov1970conformal}%
  \BibitemOpen
  \bibfield  {author} {\bibinfo {author} {\bibfnamefont {Alexander~M}\
  \bibnamefont {Polyakov}},\ }\bibfield  {title} {\enquote {\bibinfo {title}
  {Conformal symmetry of critical fluctuations},}\ }\href@noop {} {\bibfield
  {journal} {\bibinfo  {journal} {JETP Lett.}\ }\textbf {\bibinfo {volume}
  {12}},\ \bibinfo {pages} {381--383} (\bibinfo {year} {1970})}\BibitemShut
  {NoStop}%
\bibitem [{\citenamefont {Onsager}(1944)}]{Onsager1944}%
  \BibitemOpen
  \bibfield  {author} {\bibinfo {author} {\bibfnamefont {Lars}\ \bibnamefont
  {Onsager}},\ }\bibfield  {title} {\enquote {\bibinfo {title} {Crystal
  statistics. i. a two-dimensional model with an order-disorder transition},}\
  }\href {\doibase 10.1103/PhysRev.65.117} {\bibfield  {journal} {\bibinfo
  {journal} {Phys. Rev.}\ }\textbf {\bibinfo {volume} {65}},\ \bibinfo {pages}
  {117--149} (\bibinfo {year} {1944})}\BibitemShut {NoStop}%
\bibitem [{\citenamefont {Belavin}\ \emph {et~al.}(1984)\citenamefont
  {Belavin}, \citenamefont {Polyakov},\ and\ \citenamefont
  {Zamolodchikov}}]{Belavin1984}%
  \BibitemOpen
  \bibfield  {author} {\bibinfo {author} {\bibfnamefont {A.A.}\ \bibnamefont
  {Belavin}}, \bibinfo {author} {\bibfnamefont {A.M.}\ \bibnamefont
  {Polyakov}}, \ and\ \bibinfo {author} {\bibfnamefont {A.B.}\ \bibnamefont
  {Zamolodchikov}},\ }\bibfield  {title} {\enquote {\bibinfo {title} {Infinite
  conformal symmetry in two-dimensional quantum field theory},}\ }\href
  {\doibase https://doi.org/10.1016/0550-3213(84)90052-X} {\bibfield  {journal}
  {\bibinfo  {journal} {Nuclear Physics B}\ }\textbf {\bibinfo {volume}
  {241}},\ \bibinfo {pages} {333--380} (\bibinfo {year} {1984})}\BibitemShut
  {NoStop}%
\bibitem [{\citenamefont {Polchinski}(1988)}]{polchinski1988scale}%
  \BibitemOpen
  \bibfield  {author} {\bibinfo {author} {\bibfnamefont {Joseph}\ \bibnamefont
  {Polchinski}},\ }\bibfield  {title} {\enquote {\bibinfo {title} {Scale and
  conformal invariance in quantum field theory},}\ }\href@noop {} {\bibfield
  {journal} {\bibinfo  {journal} {Nuclear Physics B}\ }\textbf {\bibinfo
  {volume} {303}},\ \bibinfo {pages} {226--236} (\bibinfo {year}
  {1988})}\BibitemShut {NoStop}%
\bibitem [{\citenamefont {{Dymarsky}}\ \emph {et~al.}(2015)\citenamefont
  {{Dymarsky}}, \citenamefont {{Komargodski}}, \citenamefont {{Schwimmer}},\
  and\ \citenamefont {{Theisen}}}]{Dymarsky2015scale}%
  \BibitemOpen
  \bibfield  {author} {\bibinfo {author} {\bibfnamefont {Anatoly}\ \bibnamefont
  {{Dymarsky}}}, \bibinfo {author} {\bibfnamefont {Zohar}\ \bibnamefont
  {{Komargodski}}}, \bibinfo {author} {\bibfnamefont {Adam}\ \bibnamefont
  {{Schwimmer}}}, \ and\ \bibinfo {author} {\bibfnamefont {Stefan}\
  \bibnamefont {{Theisen}}},\ }\bibfield  {title} {\enquote {\bibinfo {title}
  {{On scale and conformal invariance in four dimensions}},}\ }\href {\doibase
  10.1007/JHEP10(2015)171} {\bibfield  {journal} {\bibinfo  {journal} {Journal
  of High Energy Physics}\ }\textbf {\bibinfo {volume} {2015}},\ \bibinfo {eid}
  {171} (\bibinfo {year} {2015})},\ \Eprint {http://arxiv.org/abs/1309.2921}
  {arXiv:1309.2921 [hep-th]} \BibitemShut {NoStop}%
\bibitem [{\citenamefont {Philippe~Francesco}(1997)}]{yellowbook}%
  \BibitemOpen
  \bibfield  {author} {\bibinfo {author} {\bibfnamefont {David~Sénéchal}\
  \bibnamefont {Philippe~Francesco}, \bibfnamefont {Pierre~Mathieu}},\
  }\href@noop {} {\emph {\bibinfo {title} {Conformal Field Theory}}},\ Graduate
  Texts in Contemporary Physics\ (\bibinfo  {publisher} {Springer New York,
  NY},\ \bibinfo {year} {1997})\BibitemShut {NoStop}%
\bibitem [{\citenamefont {Cardy}(1984)}]{Cardy1984}%
  \BibitemOpen
  \bibfield  {author} {\bibinfo {author} {\bibfnamefont {J~L}\ \bibnamefont
  {Cardy}},\ }\bibfield  {title} {\enquote {\bibinfo {title} {Conformal
  invariance and universality in finite-size scaling},}\ }\href {\doibase
  10.1088/0305-4470/17/7/003} {\bibfield  {journal} {\bibinfo  {journal}
  {Journal of Physics A: Mathematical and General}\ }\textbf {\bibinfo {volume}
  {17}},\ \bibinfo {pages} {L385--L387} (\bibinfo {year} {1984})}\BibitemShut
  {NoStop}%
\bibitem [{\citenamefont {Weigel}\ and\ \citenamefont
  {Janke}(2000)}]{Weigel2000}%
  \BibitemOpen
  \bibfield  {author} {\bibinfo {author} {\bibfnamefont {M}~\bibnamefont
  {Weigel}}\ and\ \bibinfo {author} {\bibfnamefont {W}~\bibnamefont {Janke}},\
  }\bibfield  {title} {\enquote {\bibinfo {title} {Universal amplitude-exponent
  relation for the ising model on sphere-like lattices},}\ }\href {\doibase
  10.1209/epl/i2000-00377-0} {\bibfield  {journal} {\bibinfo  {journal}
  {Europhysics Letters ({EPL})}\ }\textbf {\bibinfo {volume} {51}},\ \bibinfo
  {pages} {578--583} (\bibinfo {year} {2000})}\BibitemShut {NoStop}%
\bibitem [{\citenamefont {Deng}\ and\ \citenamefont
  {Bl\"ote}(2002)}]{Deng2002Conformal}%
  \BibitemOpen
  \bibfield  {author} {\bibinfo {author} {\bibfnamefont {Youjin}\ \bibnamefont
  {Deng}}\ and\ \bibinfo {author} {\bibfnamefont {Henk W.~J.}\ \bibnamefont
  {Bl\"ote}},\ }\bibfield  {title} {\enquote {\bibinfo {title} {Conformal
  invariance of the ising model in three dimensions},}\ }\href {\doibase
  10.1103/PhysRevLett.88.190602} {\bibfield  {journal} {\bibinfo  {journal}
  {Phys. Rev. Lett.}\ }\textbf {\bibinfo {volume} {88}},\ \bibinfo {pages}
  {190602} (\bibinfo {year} {2002})}\BibitemShut {NoStop}%
\bibitem [{\citenamefont {{Bill{\'o}}}\ \emph {et~al.}(2013)\citenamefont
  {{Bill{\'o}}}, \citenamefont {{Caselle}}, \citenamefont {{Gaiotto}},
  \citenamefont {{Gliozzi}}, \citenamefont {{Meineri}},\ and\ \citenamefont
  {{Pellegrini}}}]{Billo2013Line}%
  \BibitemOpen
  \bibfield  {author} {\bibinfo {author} {\bibfnamefont {M.}~\bibnamefont
  {{Bill{\'o}}}}, \bibinfo {author} {\bibfnamefont {M.}~\bibnamefont
  {{Caselle}}}, \bibinfo {author} {\bibfnamefont {D.}~\bibnamefont
  {{Gaiotto}}}, \bibinfo {author} {\bibfnamefont {F.}~\bibnamefont
  {{Gliozzi}}}, \bibinfo {author} {\bibfnamefont {M.}~\bibnamefont
  {{Meineri}}}, \ and\ \bibinfo {author} {\bibfnamefont {R.}~\bibnamefont
  {{Pellegrini}}},\ }\bibfield  {title} {\enquote {\bibinfo {title} {{Line
  defects in the 3d Ising model}},}\ }\href@noop {} {\bibfield  {journal}
  {\bibinfo  {journal} {arXiv e-prints}\ ,\ \bibinfo {eid} {arXiv:1304.4110}}
  (\bibinfo {year} {2013})},\ \Eprint {http://arxiv.org/abs/1304.4110}
  {arXiv:1304.4110 [hep-th]} \BibitemShut {NoStop}%
\bibitem [{\citenamefont {{Cosme}}\ \emph {et~al.}(2015)\citenamefont
  {{Cosme}}, \citenamefont {{Lopes}},\ and\ \citenamefont
  {{Penedones}}}]{Cosme2015Sphere}%
  \BibitemOpen
  \bibfield  {author} {\bibinfo {author} {\bibfnamefont {Catarina}\
  \bibnamefont {{Cosme}}}, \bibinfo {author} {\bibfnamefont {J.~M.
  Viana~Parente}\ \bibnamefont {{Lopes}}}, \ and\ \bibinfo {author}
  {\bibfnamefont {Jo{\~a}o}\ \bibnamefont {{Penedones}}},\ }\bibfield  {title}
  {\enquote {\bibinfo {title} {{Conformal symmetry of the critical 3D Ising
  model inside a sphere}},}\ }\href {\doibase 10.1007/JHEP08(2015)022}
  {\bibfield  {journal} {\bibinfo  {journal} {Journal of High Energy Physics}\
  }\textbf {\bibinfo {volume} {2015}},\ \bibinfo {eid} {22} (\bibinfo {year}
  {2015})},\ \Eprint {http://arxiv.org/abs/1503.02011} {arXiv:1503.02011
  [hep-th]} \BibitemShut {NoStop}%
\bibitem [{\citenamefont {{Schuler}}\ \emph {et~al.}(2016)\citenamefont
  {{Schuler}}, \citenamefont {{Whitsitt}}, \citenamefont {{Henry}},
  \citenamefont {{Sachdev}},\ and\ \citenamefont
  {{L{\"a}uchli}}}]{Schuler2016Universal}%
  \BibitemOpen
  \bibfield  {author} {\bibinfo {author} {\bibfnamefont {Michael}\ \bibnamefont
  {{Schuler}}}, \bibinfo {author} {\bibfnamefont {Seth}\ \bibnamefont
  {{Whitsitt}}}, \bibinfo {author} {\bibfnamefont {Louis-Paul}\ \bibnamefont
  {{Henry}}}, \bibinfo {author} {\bibfnamefont {Subir}\ \bibnamefont
  {{Sachdev}}}, \ and\ \bibinfo {author} {\bibfnamefont {Andreas~M.}\
  \bibnamefont {{L{\"a}uchli}}},\ }\bibfield  {title} {\enquote {\bibinfo
  {title} {{Universal Signatures of Quantum Critical Points from Finite-Size
  Torus Spectra: A Window into the Operator Content of Higher-Dimensional
  Conformal Field Theories}},}\ }\href {\doibase
  10.1103/PhysRevLett.117.210401} {\bibfield  {journal} {\bibinfo  {journal}
  {\prl}\ }\textbf {\bibinfo {volume} {117}},\ \bibinfo {eid} {210401}
  (\bibinfo {year} {2016})},\ \Eprint {http://arxiv.org/abs/1603.03042}
  {arXiv:1603.03042 [cond-mat.str-el]} \BibitemShut {NoStop}%
\bibitem [{\citenamefont {{Meneses}}\ \emph {et~al.}(2019)\citenamefont
  {{Meneses}}, \citenamefont {{Penedones}}, \citenamefont {{Rychkov}},
  \citenamefont {{Viana Parente Lopes}},\ and\ \citenamefont
  {{Yvernay}}}]{Meneses2019viral}%
  \BibitemOpen
  \bibfield  {author} {\bibinfo {author} {\bibfnamefont {Sim{\~a}o}\
  \bibnamefont {{Meneses}}}, \bibinfo {author} {\bibfnamefont {Jo{\~a}o}\
  \bibnamefont {{Penedones}}}, \bibinfo {author} {\bibfnamefont {Slava}\
  \bibnamefont {{Rychkov}}}, \bibinfo {author} {\bibfnamefont {J.~M.}\
  \bibnamefont {{Viana Parente Lopes}}}, \ and\ \bibinfo {author}
  {\bibfnamefont {Pierre}\ \bibnamefont {{Yvernay}}},\ }\bibfield  {title}
  {\enquote {\bibinfo {title} {{A structural test for the conformal invariance
  of the critical 3d Ising model}},}\ }\href {\doibase 10.1007/JHEP04(2019)115}
  {\bibfield  {journal} {\bibinfo  {journal} {Journal of High Energy Physics}\
  }\textbf {\bibinfo {volume} {2019}},\ \bibinfo {eid} {115} (\bibinfo {year}
  {2019})},\ \Eprint {http://arxiv.org/abs/1802.02319} {arXiv:1802.02319
  [hep-th]} \BibitemShut {NoStop}%
\bibitem [{\citenamefont {Cardy}(1985)}]{CARDY1985}%
  \BibitemOpen
  \bibfield  {author} {\bibinfo {author} {\bibfnamefont {J~L}\ \bibnamefont
  {Cardy}},\ }\bibfield  {title} {\enquote {\bibinfo {title} {Universal
  amplitudes in finite-size scaling: generalisation to arbitrary
  dimensionality},}\ }\href {\doibase 10.1088/0305-4470/18/13/005} {\bibfield
  {journal} {\bibinfo  {journal} {Journal of Physics A: Mathematical and
  General}\ }\textbf {\bibinfo {volume} {18}},\ \bibinfo {pages} {L757--L760}
  (\bibinfo {year} {1985})}\BibitemShut {NoStop}%
\bibitem [{\citenamefont {Bl\"ote}\ \emph {et~al.}(1986)\citenamefont
  {Bl\"ote}, \citenamefont {Cardy},\ and\ \citenamefont
  {Nightingale}}]{Blote1986Conformal}%
  \BibitemOpen
  \bibfield  {author} {\bibinfo {author} {\bibfnamefont {H.~W.~J.}\
  \bibnamefont {Bl\"ote}}, \bibinfo {author} {\bibfnamefont {John~L.}\
  \bibnamefont {Cardy}}, \ and\ \bibinfo {author} {\bibfnamefont {M.~P.}\
  \bibnamefont {Nightingale}},\ }\bibfield  {title} {\enquote {\bibinfo {title}
  {Conformal invariance, the central charge, and universal finite-size
  amplitudes at criticality},}\ }\href {\doibase 10.1103/PhysRevLett.56.742}
  {\bibfield  {journal} {\bibinfo  {journal} {Phys. Rev. Lett.}\ }\textbf
  {\bibinfo {volume} {56}},\ \bibinfo {pages} {742--745} (\bibinfo {year}
  {1986})}\BibitemShut {NoStop}%
\bibitem [{\citenamefont {Affleck}(1988)}]{affleck1988universal}%
  \BibitemOpen
  \bibfield  {author} {\bibinfo {author} {\bibfnamefont {Ian}\ \bibnamefont
  {Affleck}},\ }\bibfield  {title} {\enquote {\bibinfo {title} {Universal term
  in the free energy at a critical point and the conformal anomaly},}\ }in\
  \href@noop {} {\emph {\bibinfo {booktitle} {Current Physics--Sources and
  Comments}}},\ Vol.~\bibinfo {volume} {2}\ (\bibinfo  {publisher} {Elsevier},\
  \bibinfo {year} {1988})\ pp.\ \bibinfo {pages} {347--349}\BibitemShut
  {NoStop}%
\bibitem [{\citenamefont {{Milsted}}\ and\ \citenamefont
  {{Vidal}}(2017)}]{Milsted2017}%
  \BibitemOpen
  \bibfield  {author} {\bibinfo {author} {\bibfnamefont {Ashley}\ \bibnamefont
  {{Milsted}}}\ and\ \bibinfo {author} {\bibfnamefont {Guifre}\ \bibnamefont
  {{Vidal}}},\ }\bibfield  {title} {\enquote {\bibinfo {title} {{Extraction of
  conformal data in critical quantum spin chains using the Koo-Saleur
  formula}},}\ }\href {\doibase 10.1103/PhysRevB.96.245105} {\bibfield
  {journal} {\bibinfo  {journal} {\prb}\ }\textbf {\bibinfo {volume} {96}},\
  \bibinfo {eid} {245105} (\bibinfo {year} {2017})},\ \Eprint
  {http://arxiv.org/abs/1706.01436} {arXiv:1706.01436 [cond-mat.str-el]}
  \BibitemShut {NoStop}%
\bibitem [{\citenamefont {Zou}\ \emph {et~al.}(2018)\citenamefont {Zou},
  \citenamefont {Milsted},\ and\ \citenamefont {Vidal}}]{Zou2018}%
  \BibitemOpen
  \bibfield  {author} {\bibinfo {author} {\bibfnamefont {Yijian}\ \bibnamefont
  {Zou}}, \bibinfo {author} {\bibfnamefont {Ashley}\ \bibnamefont {Milsted}}, \
  and\ \bibinfo {author} {\bibfnamefont {Guifre}\ \bibnamefont {Vidal}},\
  }\bibfield  {title} {\enquote {\bibinfo {title} {Conformal data and
  renormalization group flow in critical quantum spin chains using periodic
  uniform matrix product states},}\ }\href {\doibase
  10.1103/PhysRevLett.121.230402} {\bibfield  {journal} {\bibinfo  {journal}
  {Phys. Rev. Lett.}\ }\textbf {\bibinfo {volume} {121}},\ \bibinfo {pages}
  {230402} (\bibinfo {year} {2018})}\BibitemShut {NoStop}%
\bibitem [{\citenamefont {{Brower}}\ \emph {et~al.}(2013)\citenamefont
  {{Brower}}, \citenamefont {{Fleming}},\ and\ \citenamefont
  {{Neuberger}}}]{Brower2013Lattice}%
  \BibitemOpen
  \bibfield  {author} {\bibinfo {author} {\bibfnamefont {R.~C.}\ \bibnamefont
  {{Brower}}}, \bibinfo {author} {\bibfnamefont {G.~T.}\ \bibnamefont
  {{Fleming}}}, \ and\ \bibinfo {author} {\bibfnamefont {H.}~\bibnamefont
  {{Neuberger}}},\ }\bibfield  {title} {\enquote {\bibinfo {title} {{Lattice
  radial quantization: 3D Ising}},}\ }\href {\doibase
  10.1016/j.physletb.2013.03.009} {\bibfield  {journal} {\bibinfo  {journal}
  {Physics Letters B}\ }\textbf {\bibinfo {volume} {721}},\ \bibinfo {pages}
  {299--305} (\bibinfo {year} {2013})},\ \Eprint
  {http://arxiv.org/abs/1212.6190} {arXiv:1212.6190 [hep-lat]} \BibitemShut
  {NoStop}%
\bibitem [{\citenamefont {{Brower}}\ \emph {et~al.}(2021)\citenamefont
  {{Brower}}, \citenamefont {{Fleming}}, \citenamefont {{Gasbarro}},
  \citenamefont {{Howarth}}, \citenamefont {{Raben}}, \citenamefont {{Tan}},\
  and\ \citenamefont {{Weinberg}}}]{Brower2021Radial}%
  \BibitemOpen
  \bibfield  {author} {\bibinfo {author} {\bibfnamefont {Richard~C.}\
  \bibnamefont {{Brower}}}, \bibinfo {author} {\bibfnamefont {George~T.}\
  \bibnamefont {{Fleming}}}, \bibinfo {author} {\bibfnamefont {Andrew~D.}\
  \bibnamefont {{Gasbarro}}}, \bibinfo {author} {\bibfnamefont {Dean}\
  \bibnamefont {{Howarth}}}, \bibinfo {author} {\bibfnamefont {Timothy~G.}\
  \bibnamefont {{Raben}}}, \bibinfo {author} {\bibfnamefont {Chung-I.}\
  \bibnamefont {{Tan}}}, \ and\ \bibinfo {author} {\bibfnamefont {Evan~S.}\
  \bibnamefont {{Weinberg}}},\ }\bibfield  {title} {\enquote {\bibinfo {title}
  {{Radial lattice quantization of 3D {\ensuremath{\phi}}$^{4}$ field
  theory}},}\ }\href {\doibase 10.1103/PhysRevD.104.094502} {\bibfield
  {journal} {\bibinfo  {journal} {\prd}\ }\textbf {\bibinfo {volume} {104}},\
  \bibinfo {eid} {094502} (\bibinfo {year} {2021})},\ \Eprint
  {http://arxiv.org/abs/2006.15636} {arXiv:2006.15636 [hep-lat]} \BibitemShut
  {NoStop}%
\bibitem [{\citenamefont {Madore}(1992)}]{madore1992fuzzy}%
  \BibitemOpen
  \bibfield  {author} {\bibinfo {author} {\bibfnamefont {John}\ \bibnamefont
  {Madore}},\ }\bibfield  {title} {\enquote {\bibinfo {title} {The fuzzy
  sphere},}\ }\href@noop {} {\bibfield  {journal} {\bibinfo  {journal}
  {Classical and Quantum Gravity}\ }\textbf {\bibinfo {volume} {9}},\ \bibinfo
  {pages} {69} (\bibinfo {year} {1992})}\BibitemShut {NoStop}%
\bibitem [{\citenamefont {Ippoliti}\ \emph {et~al.}(2018)\citenamefont
  {Ippoliti}, \citenamefont {Mong}, \citenamefont {Assaad},\ and\ \citenamefont
  {Zaletel}}]{Ippoliti2018Half}%
  \BibitemOpen
  \bibfield  {author} {\bibinfo {author} {\bibfnamefont {Matteo}\ \bibnamefont
  {Ippoliti}}, \bibinfo {author} {\bibfnamefont {Roger S.~K.}\ \bibnamefont
  {Mong}}, \bibinfo {author} {\bibfnamefont {Fakher~F.}\ \bibnamefont
  {Assaad}}, \ and\ \bibinfo {author} {\bibfnamefont {Michael~P.}\ \bibnamefont
  {Zaletel}},\ }\bibfield  {title} {\enquote {\bibinfo {title} {Half-filled
  landau levels: A continuum and sign-free regularization for three-dimensional
  quantum critical points},}\ }\href {\doibase 10.1103/PhysRevB.98.235108}
  {\bibfield  {journal} {\bibinfo  {journal} {Phys. Rev. B}\ }\textbf {\bibinfo
  {volume} {98}},\ \bibinfo {pages} {235108} (\bibinfo {year}
  {2018})}\BibitemShut {NoStop}%
\bibitem [{\citenamefont {Poland}\ \emph {et~al.}(2019)\citenamefont {Poland},
  \citenamefont {Rychkov},\ and\ \citenamefont {Vichi}}]{RMP_CB}%
  \BibitemOpen
  \bibfield  {author} {\bibinfo {author} {\bibfnamefont {David}\ \bibnamefont
  {Poland}}, \bibinfo {author} {\bibfnamefont {Slava}\ \bibnamefont {Rychkov}},
  \ and\ \bibinfo {author} {\bibfnamefont {Alessandro}\ \bibnamefont {Vichi}},\
  }\bibfield  {title} {\enquote {\bibinfo {title} {The conformal bootstrap:
  Theory, numerical techniques, and applications},}\ }\href {\doibase
  10.1103/RevModPhys.91.015002} {\bibfield  {journal} {\bibinfo  {journal}
  {Rev. Mod. Phys.}\ }\textbf {\bibinfo {volume} {91}},\ \bibinfo {pages}
  {015002} (\bibinfo {year} {2019})}\BibitemShut {NoStop}%
\bibitem [{\citenamefont {Simmons-Duffin}(2017)}]{Ising_CB}%
  \BibitemOpen
  \bibfield  {author} {\bibinfo {author} {\bibfnamefont {D.}~\bibnamefont
  {Simmons-Duffin}},\ }\bibfield  {title} {\enquote {\bibinfo {title} {The
  lightcone bootstrap and the spectrum of the 3d ising cft},}\ }\href
  {https://doi.org/10.1007/JHEP03(2017)086} {\bibfield  {journal} {\bibinfo
  {journal} {J. High Energ. Phys.}\ }\textbf {\bibinfo {volume} {2017}},\
  \bibinfo {pages} {86} (\bibinfo {year} {2017})}\BibitemShut {NoStop}%
\bibitem [{\citenamefont {Rychkov}\ and\ \citenamefont
  {Vichi}(2009)}]{Rychkov:2009ij}%
  \BibitemOpen
  \bibfield  {author} {\bibinfo {author} {\bibfnamefont {Vyacheslav~S.}\
  \bibnamefont {Rychkov}}\ and\ \bibinfo {author} {\bibfnamefont {Alessandro}\
  \bibnamefont {Vichi}},\ }\bibfield  {title} {\enquote {\bibinfo {title}
  {{Universal Constraints on Conformal Operator Dimensions}},}\ }\href
  {\doibase 10.1103/PhysRevD.80.045006} {\bibfield  {journal} {\bibinfo
  {journal} {Phys. Rev. D}\ }\textbf {\bibinfo {volume} {80}},\ \bibinfo
  {pages} {045006} (\bibinfo {year} {2009})},\ \Eprint
  {http://arxiv.org/abs/0905.2211} {arXiv:0905.2211 [hep-th]} \BibitemShut
  {NoStop}%
\bibitem [{\citenamefont {El-Showk}\ \emph {et~al.}(2012)\citenamefont
  {El-Showk}, \citenamefont {Paulos}, \citenamefont {Poland}, \citenamefont
  {Rychkov}, \citenamefont {Simmons-Duffin},\ and\ \citenamefont
  {Vichi}}]{ElShowk:2012ht}%
  \BibitemOpen
  \bibfield  {author} {\bibinfo {author} {\bibfnamefont {Sheer}\ \bibnamefont
  {El-Showk}}, \bibinfo {author} {\bibfnamefont {Miguel~F.}\ \bibnamefont
  {Paulos}}, \bibinfo {author} {\bibfnamefont {David}\ \bibnamefont {Poland}},
  \bibinfo {author} {\bibfnamefont {Slava}\ \bibnamefont {Rychkov}}, \bibinfo
  {author} {\bibfnamefont {David}\ \bibnamefont {Simmons-Duffin}}, \ and\
  \bibinfo {author} {\bibfnamefont {Alessandro}\ \bibnamefont {Vichi}},\
  }\bibfield  {title} {\enquote {\bibinfo {title} {{Solving the 3D Ising Model
  with the Conformal Bootstrap}},}\ }\href {\doibase
  10.1103/PhysRevD.86.025022} {\bibfield  {journal} {\bibinfo  {journal} {Phys.
  Rev. D}\ }\textbf {\bibinfo {volume} {86}},\ \bibinfo {pages} {025022}
  (\bibinfo {year} {2012})},\ \Eprint {http://arxiv.org/abs/1203.6064}
  {arXiv:1203.6064 [hep-th]} \BibitemShut {NoStop}%
\bibitem [{\citenamefont {Kos}\ \emph {et~al.}(2016)\citenamefont {Kos},
  \citenamefont {Poland}, \citenamefont {Simmons-Duffin},\ and\ \citenamefont
  {Vichi}}]{Kos:2016ysd}%
  \BibitemOpen
  \bibfield  {author} {\bibinfo {author} {\bibfnamefont {Filip}\ \bibnamefont
  {Kos}}, \bibinfo {author} {\bibfnamefont {David}\ \bibnamefont {Poland}},
  \bibinfo {author} {\bibfnamefont {David}\ \bibnamefont {Simmons-Duffin}}, \
  and\ \bibinfo {author} {\bibfnamefont {Alessandro}\ \bibnamefont {Vichi}},\
  }\bibfield  {title} {\enquote {\bibinfo {title} {{Precision Islands in the
  Ising and $O(N)$ Models}},}\ }\href {\doibase 10.1007/JHEP08(2016)036}
  {\bibfield  {journal} {\bibinfo  {journal} {JHEP}\ }\textbf {\bibinfo
  {volume} {08}},\ \bibinfo {pages} {036} (\bibinfo {year} {2016})},\ \Eprint
  {http://arxiv.org/abs/1603.04436} {arXiv:1603.04436 [hep-th]} \BibitemShut
  {NoStop}%
\bibitem [{\citenamefont {Hasenbusch}(2010)}]{Hasenbusch2010}%
  \BibitemOpen
  \bibfield  {author} {\bibinfo {author} {\bibfnamefont {Martin}\ \bibnamefont
  {Hasenbusch}},\ }\bibfield  {title} {\enquote {\bibinfo {title} {Finite size
  scaling study of lattice models in the three-dimensional ising universality
  class},}\ }\href {\doibase 10.1103/PhysRevB.82.174433} {\bibfield  {journal}
  {\bibinfo  {journal} {Phys. Rev. B}\ }\textbf {\bibinfo {volume} {82}},\
  \bibinfo {pages} {174433} (\bibinfo {year} {2010})}\BibitemShut {NoStop}%
\bibitem [{\citenamefont {Ferrenberg}\ \emph {et~al.}(2018)\citenamefont
  {Ferrenberg}, \citenamefont {Xu},\ and\ \citenamefont {Landau}}]{Landau2018}%
  \BibitemOpen
  \bibfield  {author} {\bibinfo {author} {\bibfnamefont {Alan~M.}\ \bibnamefont
  {Ferrenberg}}, \bibinfo {author} {\bibfnamefont {Jiahao}\ \bibnamefont {Xu}},
  \ and\ \bibinfo {author} {\bibfnamefont {David~P.}\ \bibnamefont {Landau}},\
  }\bibfield  {title} {\enquote {\bibinfo {title} {Pushing the limits of monte
  carlo simulations for the three-dimensional ising model},}\ }\href {\doibase
  10.1103/PhysRevE.97.043301} {\bibfield  {journal} {\bibinfo  {journal} {Phys.
  Rev. E}\ }\textbf {\bibinfo {volume} {97}},\ \bibinfo {pages} {043301}
  (\bibinfo {year} {2018})}\BibitemShut {NoStop}%
\bibitem [{\citenamefont {Pelissetto}\ and\ \citenamefont
  {Vicari}(2002)}]{Vicari2002}%
  \BibitemOpen
  \bibfield  {author} {\bibinfo {author} {\bibfnamefont {Andrea}\ \bibnamefont
  {Pelissetto}}\ and\ \bibinfo {author} {\bibfnamefont {Ettore}\ \bibnamefont
  {Vicari}},\ }\bibfield  {title} {\enquote {\bibinfo {title} {Critical
  phenomena and renormalization-group theory},}\ }\href {\doibase
  https://doi.org/10.1016/S0370-1573(02)00219-3} {\bibfield  {journal}
  {\bibinfo  {journal} {Physics Reports}\ }\textbf {\bibinfo {volume} {368}},\
  \bibinfo {pages} {549--727} (\bibinfo {year} {2002})}\BibitemShut {NoStop}%
\bibitem [{\citenamefont {Delamotte}\ \emph {et~al.}(2016)\citenamefont
  {Delamotte}, \citenamefont {Tissier},\ and\ \citenamefont
  {Wschebor}}]{Wschebor2016}%
  \BibitemOpen
  \bibfield  {author} {\bibinfo {author} {\bibfnamefont {Bertrand}\
  \bibnamefont {Delamotte}}, \bibinfo {author} {\bibfnamefont {Matthieu}\
  \bibnamefont {Tissier}}, \ and\ \bibinfo {author} {\bibfnamefont {Nicol\'as}\
  \bibnamefont {Wschebor}},\ }\bibfield  {title} {\enquote {\bibinfo {title}
  {Scale invariance implies conformal invariance for the three-dimensional
  ising model},}\ }\href {\doibase 10.1103/PhysRevE.93.012144} {\bibfield
  {journal} {\bibinfo  {journal} {Phys. Rev. E}\ }\textbf {\bibinfo {volume}
  {93}},\ \bibinfo {pages} {012144} (\bibinfo {year} {2016})}\BibitemShut
  {NoStop}%
\bibitem [{\citenamefont {{Nakayama}}(2013)}]{Nakayama2013ScaleVSConformal}%
  \BibitemOpen
  \bibfield  {author} {\bibinfo {author} {\bibfnamefont {Yu}~\bibnamefont
  {{Nakayama}}},\ }\bibfield  {title} {\enquote {\bibinfo {title} {{Scale
  invariance vs conformal invariance}},}\ }\href {\doibase
  10.48550/arXiv.1302.0884} {\bibfield  {journal} {\bibinfo  {journal} {arXiv
  e-prints}\ ,\ \bibinfo {eid} {arXiv:1302.0884}} (\bibinfo {year} {2013})},\
  \Eprint {http://arxiv.org/abs/1302.0884} {arXiv:1302.0884 [hep-th]}
  \BibitemShut {NoStop}%
\bibitem [{\citenamefont {{Casini}}\ \emph {et~al.}(2011)\citenamefont
  {{Casini}}, \citenamefont {{Huerta}},\ and\ \citenamefont
  {{Myers}}}]{Casini2011Towards}%
  \BibitemOpen
  \bibfield  {author} {\bibinfo {author} {\bibfnamefont {Horacio}\ \bibnamefont
  {{Casini}}}, \bibinfo {author} {\bibfnamefont {Marina}\ \bibnamefont
  {{Huerta}}}, \ and\ \bibinfo {author} {\bibfnamefont {Robert~C.}\
  \bibnamefont {{Myers}}},\ }\bibfield  {title} {\enquote {\bibinfo {title}
  {{Towards a derivation of holographic entanglement entropy}},}\ }\href
  {\doibase 10.1007/JHEP05(2011)036} {\bibfield  {journal} {\bibinfo  {journal}
  {Journal of High Energy Physics}\ }\textbf {\bibinfo {volume} {2011}},\
  \bibinfo {eid} {36} (\bibinfo {year} {2011})},\ \Eprint
  {http://arxiv.org/abs/1102.0440} {arXiv:1102.0440 [hep-th]} \BibitemShut
  {NoStop}%
\bibitem [{\citenamefont {{Jafferis}}\ \emph {et~al.}(2011)\citenamefont
  {{Jafferis}}, \citenamefont {{Klebanov}}, \citenamefont {{Pufu}},\ and\
  \citenamefont {{Safdi}}}]{Jafferis2011}%
  \BibitemOpen
  \bibfield  {author} {\bibinfo {author} {\bibfnamefont {Daniel~L.}\
  \bibnamefont {{Jafferis}}}, \bibinfo {author} {\bibfnamefont {Igor~R.}\
  \bibnamefont {{Klebanov}}}, \bibinfo {author} {\bibfnamefont {Silviu~S.}\
  \bibnamefont {{Pufu}}}, \ and\ \bibinfo {author} {\bibfnamefont
  {Benjamin~R.}\ \bibnamefont {{Safdi}}},\ }\bibfield  {title} {\enquote
  {\bibinfo {title} {{Towards the F-theorem: mathcal\{N\} = 2 field theories on
  the three-sphere}},}\ }\href {\doibase 10.1007/JHEP06(2011)102} {\bibfield
  {journal} {\bibinfo  {journal} {Journal of High Energy Physics}\ }\textbf
  {\bibinfo {volume} {2011}},\ \bibinfo {eid} {102} (\bibinfo {year} {2011})},\
  \Eprint {http://arxiv.org/abs/1103.1181} {arXiv:1103.1181 [hep-th]}
  \BibitemShut {NoStop}%
\bibitem [{\citenamefont {{Myers}}\ and\ \citenamefont
  {{Sinha}}(2011)}]{Myers2011Holographic}%
  \BibitemOpen
  \bibfield  {author} {\bibinfo {author} {\bibfnamefont {Robert~C.}\
  \bibnamefont {{Myers}}}\ and\ \bibinfo {author} {\bibfnamefont {Aninda}\
  \bibnamefont {{Sinha}}},\ }\bibfield  {title} {\enquote {\bibinfo {title}
  {{Holographic c-theorems in arbitrary dimensions}},}\ }\href {\doibase
  10.1007/JHEP01(2011)125} {\bibfield  {journal} {\bibinfo  {journal} {Journal
  of High Energy Physics}\ }\textbf {\bibinfo {volume} {2011}},\ \bibinfo {eid}
  {125} (\bibinfo {year} {2011})},\ \Eprint {http://arxiv.org/abs/1011.5819}
  {arXiv:1011.5819 [hep-th]} \BibitemShut {NoStop}%
\bibitem [{\citenamefont {{Casini}}\ and\ \citenamefont
  {{Huerta}}(2012)}]{Casini2012Renormalization}%
  \BibitemOpen
  \bibfield  {author} {\bibinfo {author} {\bibfnamefont {H.}~\bibnamefont
  {{Casini}}}\ and\ \bibinfo {author} {\bibfnamefont {M.}~\bibnamefont
  {{Huerta}}},\ }\bibfield  {title} {\enquote {\bibinfo {title}
  {{Renormalization group running of the entanglement entropy of a circle}},}\
  }\href {\doibase 10.1103/PhysRevD.85.125016} {\bibfield  {journal} {\bibinfo
  {journal} {\prd}\ }\textbf {\bibinfo {volume} {85}},\ \bibinfo {eid} {125016}
  (\bibinfo {year} {2012})},\ \Eprint {http://arxiv.org/abs/1202.5650}
  {arXiv:1202.5650 [hep-th]} \BibitemShut {NoStop}%
\bibitem [{\citenamefont {{Berkowitz}}\ and\ \citenamefont
  {{Fleming}}(2021)}]{Berkowitz2021Binder}%
  \BibitemOpen
  \bibfield  {author} {\bibinfo {author} {\bibfnamefont {Daniel}\ \bibnamefont
  {{Berkowitz}}}\ and\ \bibinfo {author} {\bibfnamefont {George}\ \bibnamefont
  {{Fleming}}},\ }\bibfield  {title} {\enquote {\bibinfo {title} {{Critical
  Three-Dimensional Ising Model on Spheriods from the Conformal Bootstrap}},}\
  }\href@noop {} {\bibfield  {journal} {\bibinfo  {journal} {arXiv e-prints}\
  ,\ \bibinfo {eid} {arXiv:2110.12109}} (\bibinfo {year} {2021})},\ \Eprint
  {http://arxiv.org/abs/2110.12109} {arXiv:2110.12109 [hep-lat]} \BibitemShut
  {NoStop}%
\bibitem [{\citenamefont {{Rychkov}}(2016)}]{Rychkov2016lectures}%
  \BibitemOpen
  \bibfield  {author} {\bibinfo {author} {\bibfnamefont {Slava}\ \bibnamefont
  {{Rychkov}}},\ }\bibfield  {title} {\enquote {\bibinfo {title} {{EPFL
  Lectures on Conformal Field Theory in D>= 3 Dimensions}},}\ }\href@noop {}
  {\bibfield  {journal} {\bibinfo  {journal} {arXiv e-prints}\ ,\ \bibinfo
  {eid} {arXiv:1601.05000}} (\bibinfo {year} {2016})},\ \Eprint
  {http://arxiv.org/abs/1601.05000} {arXiv:1601.05000 [hep-th]} \BibitemShut
  {NoStop}%
\bibitem [{\citenamefont {Haldane}(1983)}]{Sphere_LL_Haldane}%
  \BibitemOpen
  \bibfield  {author} {\bibinfo {author} {\bibfnamefont {F.~D.~M.}\
  \bibnamefont {Haldane}},\ }\bibfield  {title} {\enquote {\bibinfo {title}
  {Fractional quantization of the hall effect: A hierarchy of incompressible
  quantum fluid states},}\ }\href {\doibase 10.1103/PhysRevLett.51.605}
  {\bibfield  {journal} {\bibinfo  {journal} {Phys. Rev. Lett.}\ }\textbf
  {\bibinfo {volume} {51}},\ \bibinfo {pages} {605--608} (\bibinfo {year}
  {1983})}\BibitemShut {NoStop}%
\bibitem [{\citenamefont {Wu}\ and\ \citenamefont
  {Yang}(1976)}]{WuYangmonopole}%
  \BibitemOpen
  \bibfield  {author} {\bibinfo {author} {\bibfnamefont {Tai~Tsun}\
  \bibnamefont {Wu}}\ and\ \bibinfo {author} {\bibfnamefont {Chen~Ning}\
  \bibnamefont {Yang}},\ }\bibfield  {title} {\enquote {\bibinfo {title} {Dirac
  monopole without strings: monopole harmonics},}\ }\href@noop {} {\bibfield
  {journal} {\bibinfo  {journal} {Nuclear Physics B}\ }\textbf {\bibinfo
  {volume} {107}},\ \bibinfo {pages} {365--380} (\bibinfo {year}
  {1976})}\BibitemShut {NoStop}%
\bibitem [{\citenamefont {{Greiter}}(2011)}]{Sphere_LL_Greiter}%
  \BibitemOpen
  \bibfield  {author} {\bibinfo {author} {\bibfnamefont {Martin}\ \bibnamefont
  {{Greiter}}},\ }\bibfield  {title} {\enquote {\bibinfo {title} {{Landau level
  quantization on the sphere}},}\ }\href {\doibase 10.1103/PhysRevB.83.115129}
  {\bibfield  {journal} {\bibinfo  {journal} {Physical Review B}\ }\textbf
  {\bibinfo {volume} {83}},\ \bibinfo {eid} {115129} (\bibinfo {year}
  {2011})},\ \Eprint {http://arxiv.org/abs/1101.3943} {arXiv:1101.3943
  [cond-mat.str-el]} \BibitemShut {NoStop}%
\bibitem [{\citenamefont {{Hasebe}}(2010)}]{Hasebe2010fuzzy}%
  \BibitemOpen
  \bibfield  {author} {\bibinfo {author} {\bibfnamefont {Kazuki}\ \bibnamefont
  {{Hasebe}}},\ }\bibfield  {title} {\enquote {\bibinfo {title} {{Hopf Maps,
  Lowest Landau Level, and Fuzzy Spheres}},}\ }\href {\doibase
  10.3842/SIGMA.2010.071} {\bibfield  {journal} {\bibinfo  {journal} {SIGMA}\
  }\textbf {\bibinfo {volume} {6}},\ \bibinfo {eid} {071} (\bibinfo {year}
  {2010})},\ \Eprint {http://arxiv.org/abs/1009.1192} {arXiv:1009.1192
  [hep-th]} \BibitemShut {NoStop}%
\bibitem [{\citenamefont {Sondhi}\ \emph {et~al.}(1993)\citenamefont {Sondhi},
  \citenamefont {Karlhede}, \citenamefont {Kivelson},\ and\ \citenamefont
  {Rezayi}}]{Sondhi1993}%
  \BibitemOpen
  \bibfield  {author} {\bibinfo {author} {\bibfnamefont {S.~L.}\ \bibnamefont
  {Sondhi}}, \bibinfo {author} {\bibfnamefont {A.}~\bibnamefont {Karlhede}},
  \bibinfo {author} {\bibfnamefont {S.~A.}\ \bibnamefont {Kivelson}}, \ and\
  \bibinfo {author} {\bibfnamefont {E.~H.}\ \bibnamefont {Rezayi}},\ }\bibfield
   {title} {\enquote {\bibinfo {title} {Skyrmions and the crossover from the
  integer to fractional quantum hall effect at small zeeman energies},}\ }\href
  {\doibase 10.1103/PhysRevB.47.16419} {\bibfield  {journal} {\bibinfo
  {journal} {Phys. Rev. B}\ }\textbf {\bibinfo {volume} {47}},\ \bibinfo
  {pages} {16419--16426} (\bibinfo {year} {1993})}\BibitemShut {NoStop}%
\bibitem [{\citenamefont {Girvin}(2000)}]{Girvin2000}%
  \BibitemOpen
  \bibfield  {author} {\bibinfo {author} {\bibfnamefont {Steven~M.}\
  \bibnamefont {Girvin}},\ }\bibfield  {title} {\enquote {\bibinfo {title}
  {Spin and isospin: Exotic order in quantum hall ferromagnets},}\ }\href
  {\doibase https://doi.org/10.1063/1.1306366} {\bibfield  {journal} {\bibinfo
  {journal} {Physics Today}\ }\textbf {\bibinfo {volume} {53}},\ \bibinfo
  {pages} {39} (\bibinfo {year} {2000})}\BibitemShut {NoStop}%
\bibitem [{\citenamefont {{Douglas}}\ and\ \citenamefont
  {{Nekrasov}}(2001)}]{noncommuQFT}%
  \BibitemOpen
  \bibfield  {author} {\bibinfo {author} {\bibfnamefont {Michael~R.}\
  \bibnamefont {{Douglas}}}\ and\ \bibinfo {author} {\bibfnamefont {Nikita~A.}\
  \bibnamefont {{Nekrasov}}},\ }\bibfield  {title} {\enquote {\bibinfo {title}
  {{Noncommutative field theory}},}\ }\href {\doibase
  10.1103/RevModPhys.73.977} {\bibfield  {journal} {\bibinfo  {journal}
  {Reviews of Modern Physics}\ }\textbf {\bibinfo {volume} {73}},\ \bibinfo
  {pages} {977--1029} (\bibinfo {year} {2001})},\ \Eprint
  {http://arxiv.org/abs/hep-th/0106048} {arXiv:hep-th/0106048 [hep-th]}
  \BibitemShut {NoStop}%
\bibitem [{\citenamefont {{Dymarsky}}\ \emph {et~al.}(2018)\citenamefont
  {{Dymarsky}}, \citenamefont {{Kos}}, \citenamefont {{Kravchuk}},
  \citenamefont {{Poland}},\ and\ \citenamefont {{Simmons-Duffin}}}]{EMT_boot}%
  \BibitemOpen
  \bibfield  {author} {\bibinfo {author} {\bibfnamefont {Anatoly}\ \bibnamefont
  {{Dymarsky}}}, \bibinfo {author} {\bibfnamefont {Filip}\ \bibnamefont
  {{Kos}}}, \bibinfo {author} {\bibfnamefont {Petr}\ \bibnamefont
  {{Kravchuk}}}, \bibinfo {author} {\bibfnamefont {David}\ \bibnamefont
  {{Poland}}}, \ and\ \bibinfo {author} {\bibfnamefont {David}\ \bibnamefont
  {{Simmons-Duffin}}},\ }\bibfield  {title} {\enquote {\bibinfo {title} {{The
  3d stress-tensor bootstrap}},}\ }\href {\doibase 10.1007/JHEP02(2018)164}
  {\bibfield  {journal} {\bibinfo  {journal} {Journal of High Energy Physics}\
  }\textbf {\bibinfo {volume} {2018}},\ \bibinfo {eid} {164} (\bibinfo {year}
  {2018})},\ \Eprint {http://arxiv.org/abs/1708.05718} {arXiv:1708.05718
  [hep-th]} \BibitemShut {NoStop}%
\bibitem [{\citenamefont {{Klebanov}}\ \emph {et~al.}(2012)\citenamefont
  {{Klebanov}}, \citenamefont {{Pufu}}, \citenamefont {{Sachdev}},\ and\
  \citenamefont {{Safdi}}}]{Klebanov2012Etanglement}%
  \BibitemOpen
  \bibfield  {author} {\bibinfo {author} {\bibfnamefont {Igor~R.}\ \bibnamefont
  {{Klebanov}}}, \bibinfo {author} {\bibfnamefont {Silviu~S.}\ \bibnamefont
  {{Pufu}}}, \bibinfo {author} {\bibfnamefont {Subir}\ \bibnamefont
  {{Sachdev}}}, \ and\ \bibinfo {author} {\bibfnamefont {Benjamin~R.}\
  \bibnamefont {{Safdi}}},\ }\bibfield  {title} {\enquote {\bibinfo {title}
  {{Entanglement entropy of 3-d conformal gauge theories with many flavors}},}\
  }\href {\doibase 10.1007/JHEP05(2012)036} {\bibfield  {journal} {\bibinfo
  {journal} {Journal of High Energy Physics}\ }\textbf {\bibinfo {volume}
  {2012}},\ \bibinfo {eid} {36} (\bibinfo {year} {2012})},\ \Eprint
  {http://arxiv.org/abs/1112.5342} {arXiv:1112.5342 [hep-th]} \BibitemShut
  {NoStop}%
\bibitem [{\citenamefont {{Casini}}\ \emph {et~al.}(2015)\citenamefont
  {{Casini}}, \citenamefont {{Huerta}}, \citenamefont {{Myers}},\ and\
  \citenamefont {{Yale}}}]{Casini2015Mutual}%
  \BibitemOpen
  \bibfield  {author} {\bibinfo {author} {\bibfnamefont {Horacio}\ \bibnamefont
  {{Casini}}}, \bibinfo {author} {\bibfnamefont {Marina}\ \bibnamefont
  {{Huerta}}}, \bibinfo {author} {\bibfnamefont {Robert~C.}\ \bibnamefont
  {{Myers}}}, \ and\ \bibinfo {author} {\bibfnamefont {Alexandre}\ \bibnamefont
  {{Yale}}},\ }\bibfield  {title} {\enquote {\bibinfo {title} {{Mutual
  information and the F-theorem}},}\ }\href {\doibase 10.1007/JHEP10(2015)003}
  {\bibfield  {journal} {\bibinfo  {journal} {Journal of High Energy Physics}\
  }\textbf {\bibinfo {volume} {2015}},\ \bibinfo {eid} {3} (\bibinfo {year}
  {2015})},\ \Eprint {http://arxiv.org/abs/1506.06195} {arXiv:1506.06195
  [hep-th]} \BibitemShut {NoStop}%
\bibitem [{\citenamefont {{Wang}}\ \emph {et~al.}(2021)\citenamefont {{Wang}},
  \citenamefont {{Zaletel}}, \citenamefont {{Mong}},\ and\ \citenamefont
  {{Assaad}}}]{Wang2021SO5WZW}%
  \BibitemOpen
  \bibfield  {author} {\bibinfo {author} {\bibfnamefont {Zhenjiu}\ \bibnamefont
  {{Wang}}}, \bibinfo {author} {\bibfnamefont {Michael~P.}\ \bibnamefont
  {{Zaletel}}}, \bibinfo {author} {\bibfnamefont {Roger S.~K.}\ \bibnamefont
  {{Mong}}}, \ and\ \bibinfo {author} {\bibfnamefont {Fakher~F.}\ \bibnamefont
  {{Assaad}}},\ }\bibfield  {title} {\enquote {\bibinfo {title} {{Phases of the
  (2 +1 ) Dimensional SO(5) Nonlinear Sigma Model with Topological Term}},}\
  }\href {\doibase 10.1103/PhysRevLett.126.045701} {\bibfield  {journal}
  {\bibinfo  {journal} {\prl}\ }\textbf {\bibinfo {volume} {126}},\ \bibinfo
  {eid} {045701} (\bibinfo {year} {2021})},\ \Eprint
  {http://arxiv.org/abs/2003.08368} {arXiv:2003.08368 [cond-mat.str-el]}
  \BibitemShut {NoStop}%
\end{thebibliography}%

\appendix
\begin{widetext}

\clearpage

\section{Haldane pseudopotential on spherical geometry}
\label{app:pseudopotential}

This section describes the expression for a general matrix element of a two-body scalar potential $V(r)$ on spherical geometry by projecting onto the lowest Landau level ($n=0$) \cite{Sphere_LL_Haldane,Sphere_LL_Greiter}. The general second-quantization form of the Hamiltonian is,
\begin{equation}\label{eq:secondquantization_H}
  H=\sum_{\sigma,\sigma'} \sum_{m_1,m_2,m_3,m_4=-s}^{s}c^{\dagger}_{m_1,\sigma} c^{\dagger}_{m_2,\sigma'} c_{m_3,\sigma'} c_{m_4,\sigma} \delta_{m_1+m_2,m_3+m_4} 
  \langle m_1,\sigma;m_2,\sigma'|V|m_3,\sigma';m_4,\sigma\rangle 
\end{equation}
where $m$ is orbital momentum and $\sigma=\uparrow,\downarrow$ is pseudospin index. Here we just take the interaction with the same pseudospin $\sigma=\sigma'$ as an example. The matrix element is given by
\begin{equation}
  \langle m_1;m_2|V|m_3;m_4\rangle = \int dr_1 \int dr_2 \Phi^*_{m_1}(r_1)\Phi^*_{m_2}(r_2)V(\mathbf r_1,\mathbf r_2)\Phi_{m_3}(r_2)\Phi_{m_4}(r_1)
\end{equation}
where $\Phi_{m}$ is  monopole harmonics functions defined in Eq. \ref{eq:orbitals}.
If we expand both the initial and final state vectors in the coupled angular
momentum basis, we can rewrite the two-body matrix element in the following form:
\begin{equation}\label{eq:V_pseudo}
   \langle m_1;m_2|V|m_3;m_4\rangle = \sum_{l,l'}\langle m_1;m_2|l,m_1+m_2\rangle \langle l',m_3+m_4|m_4;m_3\rangle  
  \langle l,m_1+m_2|V|l',m_3+m_4\rangle, 
\end{equation}
where the coefficient $\langle l,m_1+m_2|m_1;m_2\rangle $ is the Clebsch-Gordan coefficient. $\langle l,m|V|l',m'\rangle  =
V_l \delta_{l,l'}$ and the pair pseudopotential $V_l$ describes the interaction energy of a pair of electrons as a function of their pair angular momentum $l$ \cite{Sphere_LL_Haldane}.

To perform calculations in this paper, we used the second-quantization form of Hamiltonian Eq. \ref{eq:secondquantization_H} with the interaction elements shown in Eq. \ref{eq:V_pseudo}. Since we only focus on the short-ranged interactions in real-space (e.g.  $ \delta(\Omega_{ab}) , \nabla^2\delta(\Omega_{ab})$), the choice of pseudopotential are limited to $V_0,V_1$.   
The explicit relation between the pseudopotentials and the two-body interaction potential in real-space will be also presented below.

\subsection{Connection of the pseudopotential with real-space interactions}

If the potential $V$ only depends on $|\mathbf r_1- \mathbf r_2|$ on the spherical geometry,
it can be expanded in Legendre polynomials,
\begin{equation}\label{eq:Legendre_expansion}
V(|\mathbf r_1-\mathbf r_2|)= \sum_{k=0}^{\infty}U_{k}P_k(\cos\theta_{12})
\end{equation}
and the real-space interaction can be rewritten in terms of a new set of parameters $U_k$:
\begin{equation}
  U_{k}=\frac{1}{2}\int_{0}^{\pi}d\theta V(|\mathbf r_1- \mathbf r_2|)P_k(\cos\theta)\sin\theta .
\end{equation}

If we insert the potential form Eq. \ref{eq:Legendre_expansion} into the matrix element, we have 
\begin{align}
&\langle m_1;m_2|V|m_3;m_4\rangle \nonumber \\
&= 
  \int d\Omega_1 \int d\Omega_2 \Phi^{*}_{m_1}(\Omega_1)\Phi^{*}_{m_2}(\Omega_2) [\sum_{k} U_{k} \frac{1}{2k+1} \sum_{m=-k}^{k} Y^*_{km}(\Omega_1) Y_{km}(\Omega_2)] \Phi_{m_3}(\Omega_2)\Phi_{m_4}(\Omega_1) \nonumber \\
&= \sum_k U_k \frac{1}{2k+1} \sum_{m=-k}^{k} \times \int d\Omega_1 
\Phi_{m_1}^*(\Omega_1) \overline{Y}_{km}(\Omega_1)  \Phi_{m_4}(\Omega_1) \int d\Omega_2
\Phi^*_{m_2}(\Omega_2) Y_{km}(\Omega_2)  \Phi_{m_3}(\Omega_2) \nonumber\\
&= \sum_{k}   U_k(-)^{6s+m_2+m_3} (2s+1)^2 
   \left(\begin{array}{ccc}
    s & k & s \\
    -m_1 & m_1-m_3 & m_3
  \end{array}\right)
  \left(\begin{array}{ccc}
    s & k & s \\
    -m_2 & m_2-m_4 & m_4
  \end{array}\right)
    \left(\begin{array}{ccc}
    s & k & s \\
    -s & 0 & s
  \end{array}\right)^2.
\end{align}
Here, for a general Wigner 3j coefficient, $\left(\begin{array}{ccc}
	s_1 & s_2 & s_3 \\
	m_1 & m_2 & m_3
\end{array}\right)$, it
is non-zero only when $m_1 +m_2 +m_3 = 0$ and when $s_1, s_2, s_3$ together satisfy the
triangle inequality, $|s_1-s_2|\le s_3 \le s_1 + s_2$.

Through this matrix element form, one can obtain that the pseudopotential of particles in the lowest Landau level is connected with parameter $U_k$ via 
\begin{equation}\label{eq:pseudopotentialwithU}
V_{2s-l} =
   \sum_{k=0}^{2s}U_k(-)^{2s_0+l}(2s+1)^2
    \left\{\begin{array}{ccc}
    l & s & s \\
    k & s & s
  \end{array}\right\}
    \left(\begin{array}{ccc}
    s & k & s \\
    -s & 0 & s
  \end{array}\right)^2,
\end{equation}
where  $\left\{\begin{array}{ccc}
    l & s & s \\
    k & s & s
  \end{array}\right\}$ is Wigner 6j coefficient.

In this paper, we only consider the short-ranged potentials:
\begin{itemize}
	
	\item For short-ranged potential $U(\Omega_{ab}) = \delta(\Omega_{ab})$, by using the expansion  
	\begin{align}
		\delta(\Omega_a-\Omega_b) = \sum_{l=0}^{\infty} \sum_{m=-l}^{l} Y^*_{l,m}(\Omega_a) Y_{l,m}(\Omega_b) = \sum_{l=0}^{\infty} (2l+1)P_l(\cos\theta_{ab}),
	\end{align}  we have $U_l =2l+1$. 
With the help of Eq. \ref{eq:pseudopotentialwithU}, we get the pseudopotentials related to the short-ranged potential $U(\Omega_{ab}) = \delta(\Omega_{ab})$ as 
\begin{align} \label{eq:pseudoV0}
V_{2s-l}=
	\begin{cases}
			 \frac{(2s+1)^2 }{(4s+1)} , l=2s\\
			0, l\neq 2s
	\end{cases}.
 \end{align}

	\item For short-ranged potential $U(\Omega_{ab}) = \nabla^2 \delta(\Omega_{ab})$, by using the expansion  
\begin{align}
	\nabla^2_a \delta(\Omega_a-\Omega_b) =  \sum_{l=0}^{\infty} \sum_{m=-l}^{l} \nabla^2_a Y^*_{l,m}(\Omega_a) Y_{l,m}(\Omega_b) = \sum_{l=0}^{\infty} (-l(l+1))(2l+1)P_l(\cos\theta_{ab}),
\end{align}
we have $U_l = -l(l+1) (2l+1)$. With the help of Eq. \ref{eq:pseudopotentialwithU}, we get the pseudopotentials related to the short-ranged potential $U(\Omega_{ab}) = \nabla^2_a \delta(\Omega_{ab})$:  
 \begin{align}\label{eq:pseudoV1}
   V_{2s-l}=
	\begin{cases}
	 -\frac{s(2s+1)^2}{4s+1} , l=2s\\
	 \frac{s(2s+1)^2}{4s-1}	 , l=2s-1\,\,\,  \\
		0, l<2s-1
	\end{cases}.
 \end{align}
In a word, for a general two-body interaction potential on the spherical geometry $U(\Omega_{ab}) = g_0 \delta(\Omega_{ab}) +g_1\nabla^2\delta(\Omega_{ab})$, one can use
Eq. \ref{eq:pseudoV0} and \ref{eq:pseudoV1} to connect with the pseudopotentials $V_l$ as defined in Eq. \ref{eq:V_pseudo}.

\end{itemize}

\section{Physical observables across the phase transition}\label{smsec:M2}
\begin{figure}
    \centering
\includegraphics[width=0.65\textwidth]{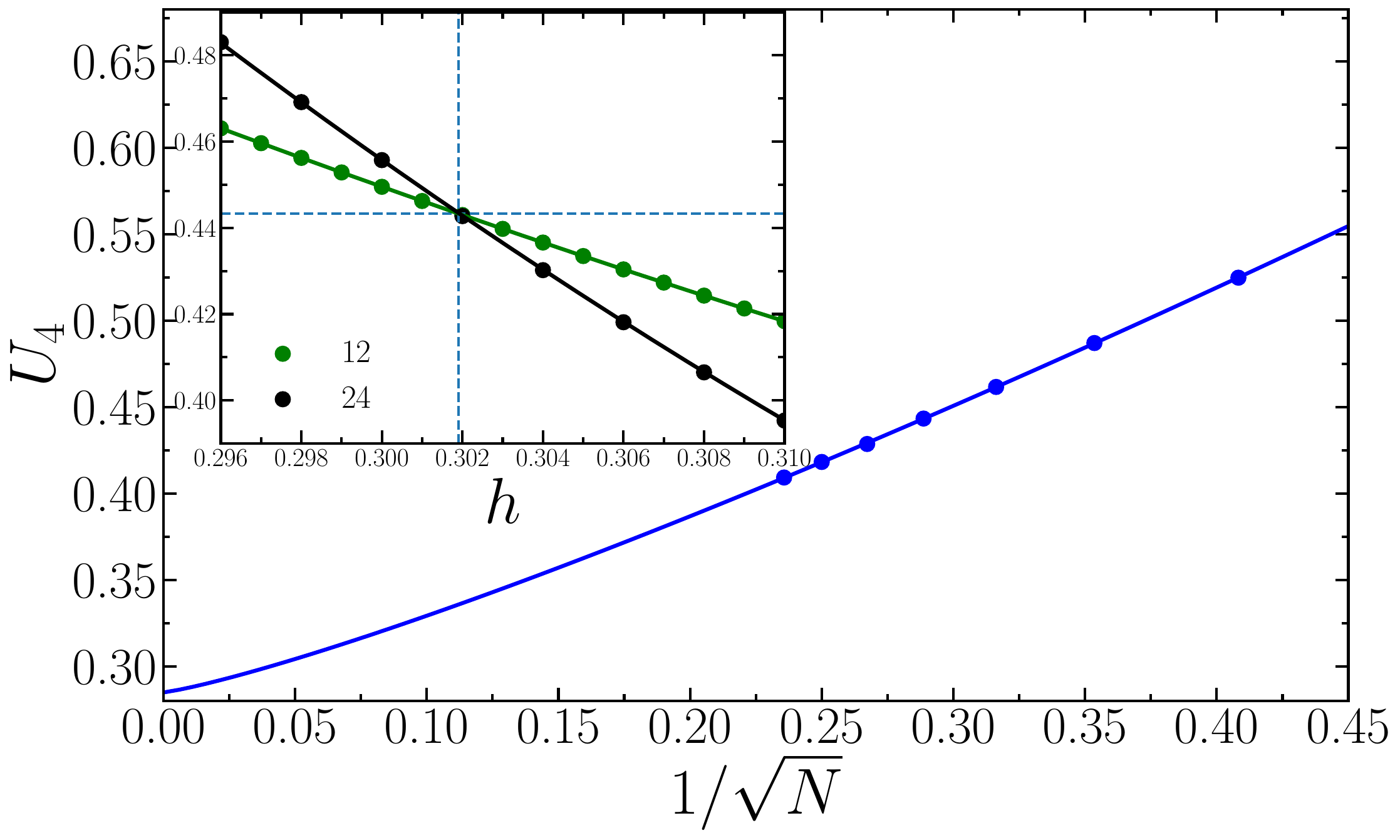}
    \caption{A
finite-size scaling analysis of the Binder cumulant $U^c_4$ at the phase transition. Each data point is determined by the crossing point on system size pair $(N, 2N)$. The analysis is according to the scaling form $U^c_4(N) =a N^{-\omega/2}+b$.  Inset: Example of finite-size crossing point $U^c_4(N)$ with $N = 12$
and $2N = 24$.}
    \label{sfig:observable}
\end{figure}

In this section, we provide more detailed analysis on the finite size scaling of physical observable $M^2$ and binder ratio $U_4$. 
In Fig. \ref{sfig:observable}, order parameter $\langle M^2\rangle$ is almost unchanged near the critical point $h\approx h_c$, which signals the phase transition point. 
In comparison, we notice that, as $N$ increases the crossing point of $U^c_4$ is less converged, which is not as perfect as the crossing of order parameter. In the finite-size scaling, the crossing value of the cumulant itself approaches its thermodynamic limit $U^c_4 \approx 0.2849 \pm 0.0063$. And we also estimate the upper bound by the lowest value that we get in the DMRG calculation. In a word, with the data up to $N=36$ the best estimate we can give is $U^c_4\sim (0.28, 0.40)$.

The larger uncertainty of binder ratio is likely due to that the $U_4$ suffers a much larger finite size effect, since at the phase transition $U_4$ is related to the four point correlator of the order parameter field $\sigma$ in CFT.  
Similar finite size effect has also been observed in Monte Carlo simulations of $3D$ classical or $2+1D$ quantum Ising transitions with much larger system size.

Additionally, we shall note that for the same universality defined or realized on distinct manifolds, $U_4$ will be generically different even in the thermodynamic limit. 
In principle $U_4$ on the conformal manifold (e.g. $R^3$, $S^2\times \mathbb{R}$) can be computed using the $R^3$ four-point correlator.~\footnote{An approximate $R^3$ four-point correlator can be reconstructed using the data from conformal bootstrap.} 
For $3D$ Ising transition on the non-conformal manifold such as $T^2\times \mathbb{R}$ or $T^3$, which Monte Carlo usually simulates, $U_4$ cannot computed using the $R^3$ four-point correlator.

\section{Excitation gap}
\subsection{Charge gap}\label{smsec:fermiongap}
In the discussion of quantum magnetism in electron systems, 
one preliminary question is if or not the charge excitation gap vanishes. Here we define the charge gap as $\Delta_c (N)= E_0(N_e+1,N)+E_0(N_e-1,N)-2E_0(N_e,N)$, where $E_0(N_e,N)$ is the ground state energy on $N$ LLL orbitals  by filling $N_e$ electrons. After obtained the charge gap on each system size, we perform a finite-size scaling to estimate the charge gap in the thermodynamic limit. 
As shown in Fig. \ref{sfig:chargegap}, the charge gap at the critical point $h=h_c$ is nonzero on all system sizes, and the value in the thermodynamic limit is also finite. Thus, we conclude that the low-energy excitation is dominated by the spin excitation other than the charge excitation.

\begin{figure}
    \centering
\includegraphics[width=0.4\textwidth]{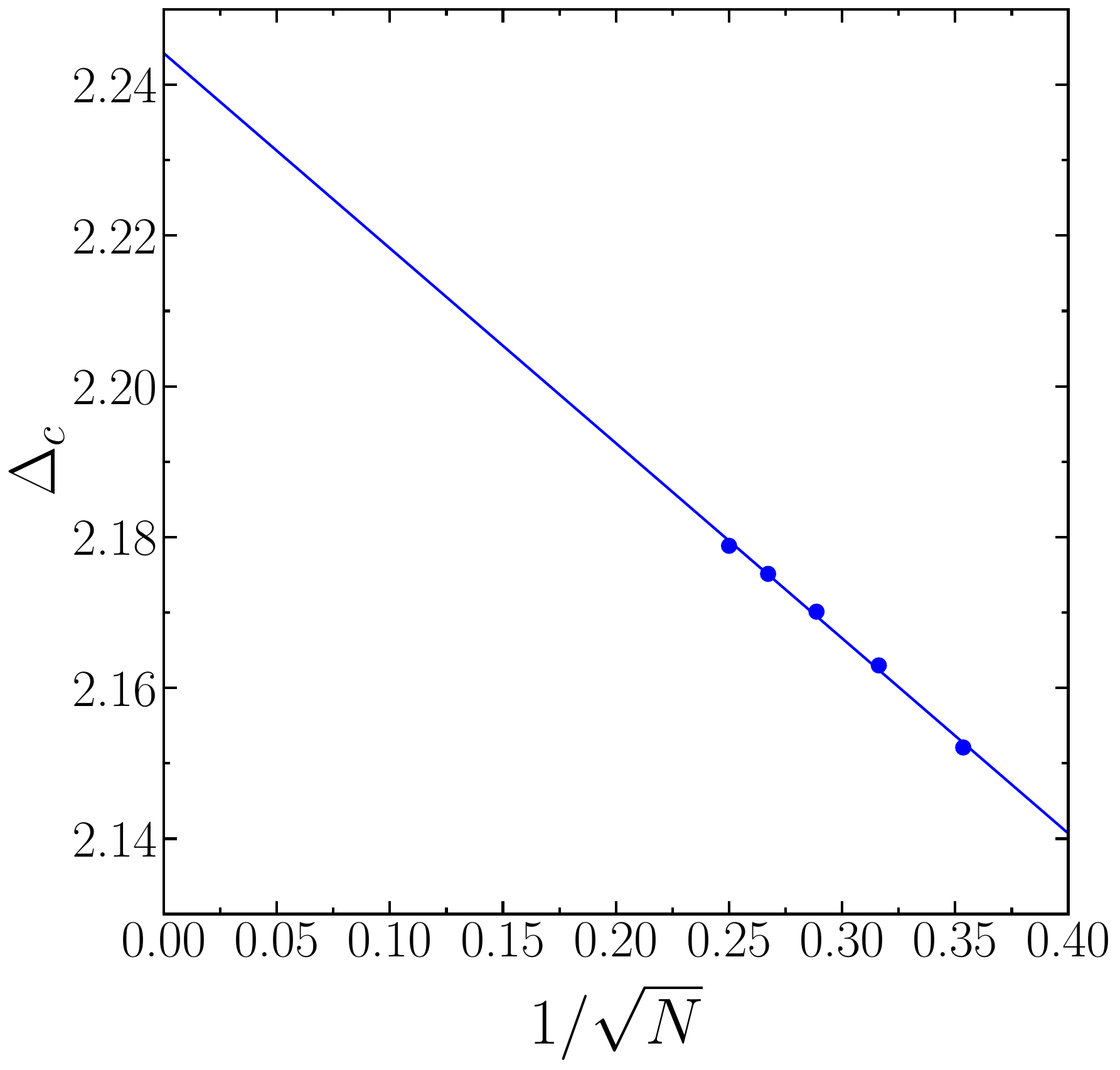}
    \caption{Finite size scaling of charge excitation gap at the phase transition. The charge gap is defined as $\Delta_c = E_0(N_e+1,N)+E_0(N_e-1,N)-2E_0(N_e,N)$, and $E_0(N_e,N)$ is the ground state energy by filling $N_e$ electrons.
    }
    \label{sfig:chargegap}
\end{figure}

\subsection{Spin excitation gap} \label{smsec:excitationgap}
In this section, we discuss the spin excitation gap. In the Ising ferromagnet ($h<h_c$), flipping a spin orientation should cost finite exchange energy, so the spin excitation gap should be nonzero. Similarly, the paramagnetic ground state $h>h_c$ is a trivial insulator, which should be separated from all other excited states by a finite energy gap. In contrast, at the critical point, the system becomes gapless, which should be distinct from the other two gapped phases. As shown in Fig. \ref{sfig:excitationgap}, we show three typical plots of excitation gap in Ising ferromagnet phase, paramagnet phase and at the phase transition point. It is clear that, the excitation gaps are finite for ferromagnet and paramagnet phase, but the system becomes gapless at the transition point $h\approx h_c$. The most interesting thing is, these critical  excitations at finite system sizes form  a characteristic conformal tower structure as discussed in the main text, which calls for a CFT description of 3D Ising criticality.  

\begin{figure}
    \centering
\includegraphics[width=0.3\textwidth]{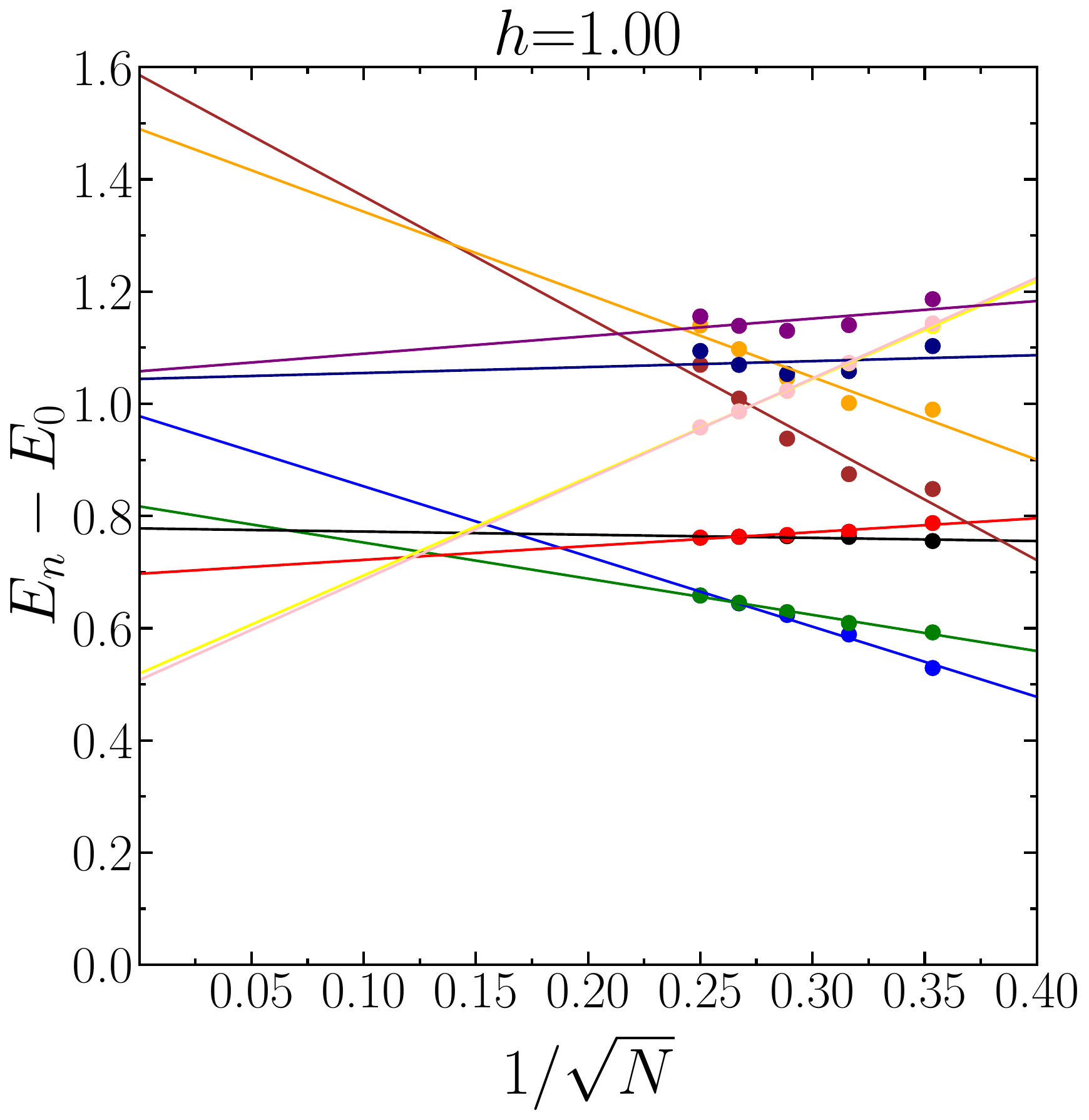}
\includegraphics[width=0.3\textwidth]{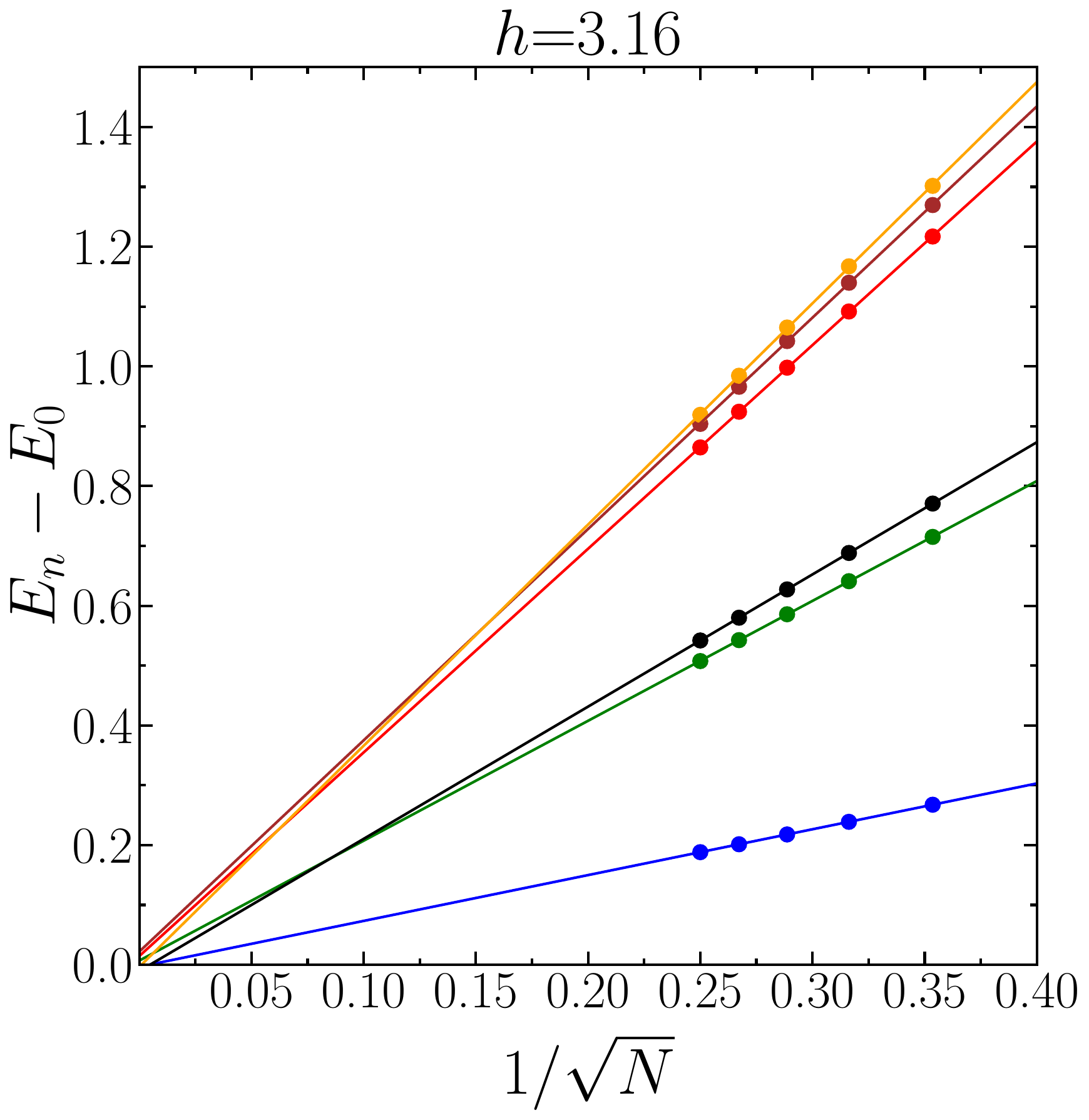}
\includegraphics[width=0.3\textwidth]{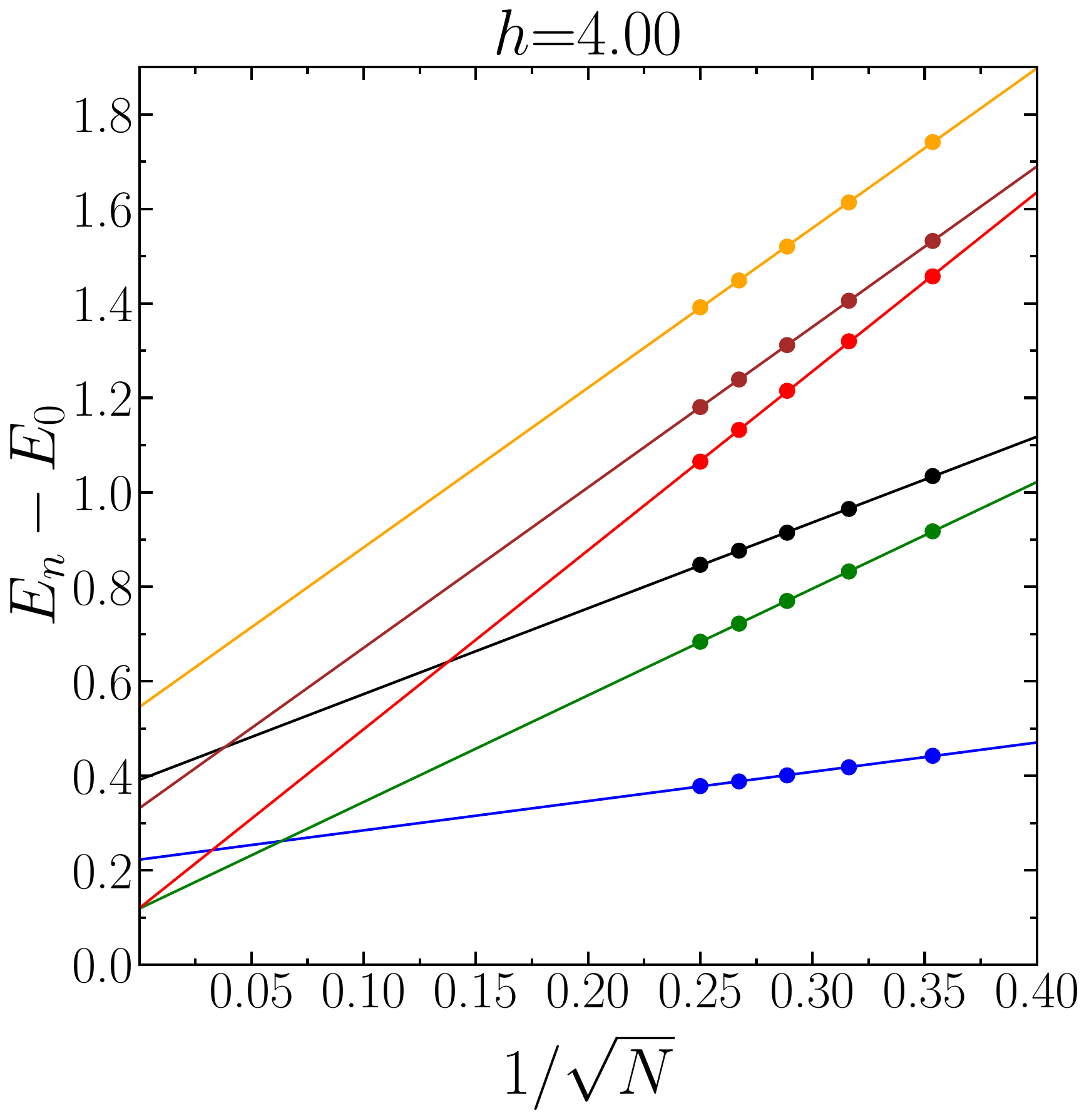}
    \caption{Finite-size scaling of the lowest six excitation gap of (a) quantum Hall ferromagnet at $h=1.0<h_c$, (b) transition point at $h=h_c$ and (c) disordered paramagnet at $h=4.0>h_c$.}
    \label{sfig:excitationgap}
\end{figure}

\section{Details of numerical data}\label{smsec:conformaldata}
In this section, we present the data of energy spectra which are organized by the good quantum numbers and conformal multiplet of various primary fields, e.g. $\epsilon$ (Tab. \ref{tab:epsilon}), $\epsilon'$ (Tab. \ref{tab:epsilonp}), $T_{\mu_1\mu_2}$ (Tab. \ref{tab:EMT}), $T_{\mu_1\mu_2}'$ (Tab. \ref{tab:EMTp}), $\epsilon_{\mu_1,\mu_2,\mu_3,\mu_4}$ (Tab. \ref{tab:epsilonL4}), $\sigma$ (Tab. \ref{tab:sigma}), $\sigma'$ (Tab. \ref{tab:sigmap}), $\sigma_{\mu_1\mu_2}$ (Tab. \ref{tab:sigmaL2}), and $\sigma_{\mu_1\mu_2\mu_3}$ (Tab. \ref{tab:sigmaL3}).
For comparison, we also list the results from conformal boostrap (CB) method \cite{Ising_CB,RMP_CB}.
These data are used for plotting Fig. \ref{fig:multiplet}. 
For the primary fields, the discrepancies are really small ( $<1.6\%$). Generally, fields with higher energies (conformal weights) have larger discrepancies, which is attributed to the finite-size effect. 
We add that these detailed data also gives a good quantification for the numerical error without inputting other results such as numerical bootstrap.
The idea is that, since the conformal symmetry predicts the integer spacings between primaries and their descendants, we can examine how good is this preserved in our spectrum.
Based on this we can give a  conservative estimate for numerical errors of primiaries and low lying descendants, which are $3\%$ relative errors.  
A rigorous error analysis based on the finite size scaling and off-critical behavior will be interesting for the future work.

Another interesting point is that, we identify almost perfect state-operator
correspondence in surprisingly small system sizes. In the main text, we only present the numerical data at a given system size, i.e. $N=16$, which is the largest system size that we can reach using ED. Here, to further elucidate that the numerical findings indeed reflect the physics in the thermodynamic limit, we show the energy spectra on different system sizes.   
In Fig. \ref{sfig:scaling_primary}, we show the energy spectra obtained on different system sizes from $N=8$ to $N=16$. 
As one can see that, the energies on \textit{all} system sizes match the prediction of 3D CFT quite well.

\begin{figure}
    \centering
\includegraphics[width=0.75\textwidth]{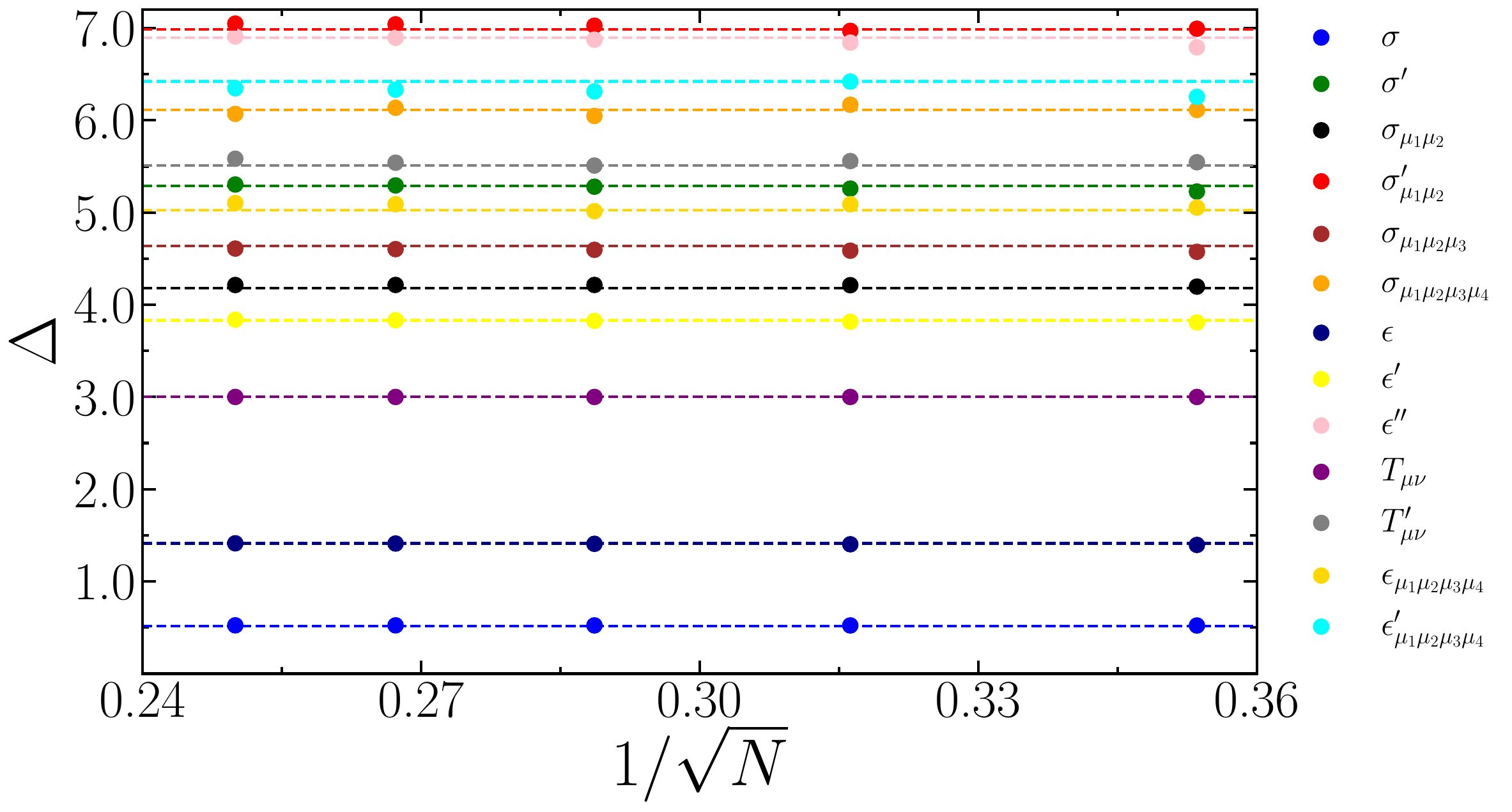}
    \caption{The energy spectra corresponding to primary fields for various system sizes $N=8-16$. 
    The dashed color lines denote the numerical values from conformal boostrap method. 
    }
    \label{sfig:scaling_primary}
\end{figure}

\begin{table}[ht]
\setlength{\tabcolsep}{0.2cm}
\renewcommand{\arraystretch}{1.4}
    \centering
    \caption{Conformal multiplet of $\epsilon$. \label{tab:epsilon}}
    \begin{tabular}{cccccc} \hline\hline
Operator & Quantum Number  & CB data & $N=16$ & Errors \\ 
$\epsilon$& $\ell=0$ & $1.412625(10)$ & 1.41355766 & 0.066\% \\ 
$\partial_\mu \epsilon$ & $\ell=1$ & $2.412625(10)$ & 2.40776449 & 0.201\% \\
$\partial_{\mu_1} \partial_{\mu_2} \epsilon$ & $\ell=2$  & $3.412625(10)$ & 3.41455749  & 0.057\% \\
$\square \epsilon$ & $\ell=0$  & $3.412625(10)$ & 3.47303235 & 1.770\% \\
$\partial_{\mu_1} \partial_{\mu_2} \partial_{\mu_3} \epsilon$ & $\ell=3$  & $4.412625(10)$ & 4.38113022 & 0.714\% \\
$\square \partial_\mu  \epsilon$ & $\ell=1$ & $4.412625(10)$ & 4.55437869 & 3.212\%  \\ 
$\partial_{\mu_1}\partial_{\mu_2} \partial_{\mu_3}\partial_{\mu_4} \epsilon$ & $\ell=4$ & $5.412625(10)$ & 5.2379631 & 3.227\% \\ 
$\partial_{\mu_1}\partial_{\mu_2} \square \epsilon$ & $\ell=2$ & $5.412625(10)$ & 5.5514904 & 2.566\% \\
$\square^2 \epsilon$ & $\ell=0$ & $5.412625(10)$ & 5.70570641 & 5.415\% \\ 
$\partial_{\mu_1} \partial_{\mu_2} \partial_{\mu_3} \square \epsilon$ & $\ell=3$ & 6.412625(10) & 6.43712303 & 0.382\% \\
$\partial_{\mu_1} \square^2 \epsilon$ &  $\ell=1$ & 6.412625(10) & 6.66423677 & 3.924\%
\\ \hline\hline
\end{tabular} \end{table}

\begin{table}[ht]
\setlength{\tabcolsep}{0.2cm}
\renewcommand{\arraystretch}{1.4}
    \centering
    \caption{Conformal mulitplet of $\epsilon'$. \label{tab:epsilonp}}
    \begin{tabular}{cccccc} \hline\hline
Operator & Quantum Number  & CB data & $N=16$ & Errors  \\ 
$\epsilon'$ & $\ell=0$ & 3.82968(23)  &  3.83772859 & 0.210\% \\ 
$\partial_{\mu_1} \epsilon'$ & $\ell=1$ & 4.82968(23)  & 4.83973617 & 0.208\%  \\ 
$\partial_{\mu_1} \partial_{\mu_2} \epsilon'$ & $\ell=2$ & 5.82968(23) & 5.82918219 & 0.009\% \\ 
$\square \epsilon'$ & $\ell=0$ & 5.82968(23)  &  5.9605325 & 2.245\% \\ 
$\partial_{\mu_1} \partial_{\mu_2}\partial_{\mu_3} \epsilon'$ & $\ell=3$ & 6.82968(23) &6.76617638 & 0.930\% \\
$\partial_{\mu_1}\square \epsilon'$ & $\ell=1$ & 6.82968(23) & 7.05458433 & 3.293\%
\\ \hline\hline
\end{tabular} \end{table}    

\begin{table}[ht]
\setlength{\tabcolsep}{0.2cm}
\renewcommand{\arraystretch}{1.4}
    \centering
    \caption{Conformal mulitplet of $T_{\mu_1\mu_2}$. \label{tab:EMT}}
    \begin{tabular}{cccccc} \hline\hline
Operator & Quantum Number  & Exact value & $N=16$ & Errors \\ 
$T_{\mu_1\mu_2}$ & $\ell=2$ & 3 & 3 & 0.000\% \\
$\partial_{\nu_1} T_{\mu_1\mu_2}$ & $\ell=3$  & 4 & 4.03219819 & 0.805\% \\ 
$ \varepsilon_{\mu_2 \rho \tau}\partial_{\rho} T_{\mu_1\mu_2}$ & $\ell=2$, $P=-1$   & 4 & 4.07392075 & 1.848\% \\
$\partial_{\nu_1} \partial_{\nu_2} T_{\mu_1\mu_2}$ & $\ell=4$ & 5 & 4.96734107 & 0.653\% \\
$ \varepsilon_{\mu_2 \rho \tau}\partial_{\rho} \partial_{\nu_1} T_{\mu_1\mu_2}$ & $\ell=3$, $P=-1$   & 5 & 5.14602926 & 2.921\% \\ 
$\square T_{\mu_1\mu_2}$ & $\ell=2$ & 5 & 5.17292963 & 3.459\% \\
$\partial_{\nu_1}\square T_{\mu_1\mu_2}$ & $\ell=3$ & 6 & 6.04586808 & 0.764\% \\ 
 $ \varepsilon_{\mu_2 \rho \tau}\partial_{\rho} \partial_{\nu_1} \partial_{\nu_2} T_{\mu_1\mu_2}$ & $\ell=4$, $P=-1$ & 6 & 6.06221026 & 1.037\% \\
$ \varepsilon_{\mu_2 \rho \tau}\partial_{\rho} \square T_{\mu_1\mu_2}$ & $\ell=2$, $P=-1$ & 6 & 6.29074558 & 4.846\%
\\ \hline\hline
\end{tabular} \end{table}

\begin{table}[ht]
\setlength{\tabcolsep}{0.2cm}
\renewcommand{\arraystretch}{1.4}
    \centering
    \caption{Conformal multiplet of $T'_{\mu_1\mu_2}$.\label{tab:EMTp}}
    \begin{tabular}{cccccc} \hline\hline
Operator & Quantum Number  & CB data & $N=16$ & Errors \\ 
$T'_{\mu_1\mu_2}$ & $\ell=2$ & 5.50915(44) & 5.5827144 & 1.335\% \\ 
$\partial_{\nu_1}T'_{\mu_1\mu_2}$ & $\ell=3$ & 6.50915(44) & 6.57137975 & 0.956\% \\ 
$ \varepsilon_{\mu_2 \rho \tau}\partial_{\rho} T'_{\mu_1\mu_2}$ & $\ell=2$, $P=-1$, & 6.50915(44) & 6.57557892 & 1.020\% \\
$\partial_{\mu_1}T'_{\mu_1\mu_2}$ & $\ell=1$ & 6.50915(44) &  6.74639599 & 3.645\%
\\ \hline\hline
\end{tabular} \end{table}

\begin{table}[ht]
\setlength{\tabcolsep}{0.2cm}
\renewcommand{\arraystretch}{1.4}
    \centering
    \caption{Conformal mulitplet of $\epsilon_{\mu_1\mu_2\mu_3\mu_4}$. \label{tab:epsilonL4}}
    \begin{tabular}{cccccc} \hline\hline
Operator & Quantum Number  & CB data & $N=16$ & Errors \\ $\epsilon_{\mu_1\mu_2\mu_3\mu_4}$ & $\ell=4$ & 5.022665(28) & 5.1029942 & 1.599\% \\ 
$\varepsilon_{\mu_4 \rho \tau}\partial_{\rho} \epsilon_{\mu_1\mu_2\mu_3\mu_4}$  & $\ell=4$, $P=-1$ & 6.022665(28) & 6.17684693 & 2.560\% \\
$\partial_{\mu_1}\epsilon_{\mu_1\mu_2\mu_3\mu_4}$ & $\ell=3$ & 6.022665(28) & 6.19439341 & 2.851\%
\\ \hline\hline
\end{tabular} \end{table}

\begin{table}[ht]
\setlength{\tabcolsep}{0.2cm}
\renewcommand{\arraystretch}{1.4}
    \centering
    \caption{Conformal multiplet of $\sigma$. \label{tab:sigma}}
    \begin{tabular}{cccccc} \hline\hline
Operator & Quantum number  & CB data & $N=16$ & Errors \\ 
$\sigma$ & $\ell=0$ & $0.5181489(10)$ & 0.52428857 & 1.185\% \\ 
$\partial_\mu \sigma$& $\ell=1$  & $1.5181489(10)$ & 1.50941793 & 0.575\% \\
$\square \sigma$ & $\ell=0$  & $2.5181489(10)$ & 2.51722181 & 0.037\% \\
$\partial_{\mu_1}\partial_{\mu_2}\sigma$ & $\ell=2$  & $2.5181489(10)$ & 2.55937503 & 1.637\% \\
$\square \partial_\mu  \sigma$ & $\ell=1$  & $3.5181489(10)$ & 3.50635346 & 0.335\% \\ 
$\partial_{\mu_1}\partial_{\mu_2}\partial_{\mu_3} \sigma$ & $\ell=3$ & $3.5181489(10)$ & 3.6059226 & 2.495\% \\
$\square \partial_{\mu_1}\partial_{\mu_2} \sigma$& $\ell=2$ & $4.5181489(10)$  & 4.47002281  & 1.065\% \\ 
$\square^2 \sigma$ & $\ell=0$  & $4.5181489(10)$ & 4.57231367 & 1.199\% \\ 
$\partial_{\mu_1}\partial_{\mu_2}\partial_{\mu_3}\partial_{\mu_4}\sigma$& $\ell=4$ & $4.5181489(10)$ & 4.52727499 & 0.202\% \\ 
$\partial_{\mu_1}\partial_{\mu_2} \partial_{\mu_3}\square \sigma$ & $\ell=3$ & 5.5181489(10) &  5.36761913 & 2.728\% \\ 
$\partial_{\mu} \square^2 \sigma$ & $\ell=1$ & 5.5181489(10) & 5.60563429 & 1.585\% \\ $\partial_{\mu_1} \partial_{\mu_2} \partial_{\mu_3} \partial_{\mu_4} \square \sigma$ & $\ell=4$ & 6.5181489(10)  & 6.24268467 & 4.226\% \\
$\partial_{\mu_1} \partial_{\mu_2}  \square^2 \sigma$ & $\ell=2$ & 6.5181489(10)  & 6.58905267 & 1.088\% \\
$\square^3 \sigma$  & $\ell=0$ &  $6.5181489(10)$ & 6.74334514 & 3.455\% 
\\ \hline\hline

\end{tabular} \end{table}

\begin{table}[ht]
\setlength{\tabcolsep}{0.2cm}
\renewcommand{\arraystretch}{1.4}
    \centering
    \caption{Conformal multiplet of $\sigma'$. \label{tab:sigmap}}
    \begin{tabular}{cccccc} \hline\hline
Operator & Quantum number  & CB data & $N=16$ & Errors  \\ 
$\sigma'$ & $\ell=0$ &  5.2906(11)  & 5.30346641 & 0.243\% \\
$\partial_{\mu_1} \sigma'$ & $\ell=1$ & 6.2906(11) & 6.27713785 & 0.214\%
\\ \hline\hline
\end{tabular} \end{table}

\begin{table}[ht]
\setlength{\tabcolsep}{0.2cm}
\renewcommand{\arraystretch}{1.4}
    \centering
    \caption{Conformal multiplet of $\sigma_{\mu_1\mu_2}$. \label{tab:sigmaL2}}
    \begin{tabular}{cccccc} \hline\hline
Operator & Quantum number  & CB data & $N=16$ & Errors  \\ 
$\sigma_{\mu_1\mu_2}$ & $\ell=2$ &  4.180305(18) & 4.21382989 & 0.802\% \\
$\partial_{\nu_1}\sigma_{\mu_1\mu_2}$ & $\ell=3$ & 5.180305(18) & 5.23649044 & 1.085\%  \\
$\partial_{\mu_1} \sigma_{\mu_1\mu_2}$ & $\ell=1$ & 5.180305(18) & 5.31575894 & 2.615\% \\ 
$\varepsilon_{\mu_2 \rho \tau}\partial_{\rho} \sigma_{\mu_1\mu_2}$ &$\ell=2$, $P=-1$ & 5.180305(18) & 5.25415317 & 1.426\% \\
$\partial_{\nu_1}\partial_{\nu_2}\sigma_{\mu_1\mu_2}$ & $\ell=4$  & 6.180305(18) & 6.18724938 & 0.112\% \\ 
$\varepsilon_{\mu_2 \rho \tau}\partial_{\rho} \partial_{\nu_1} \sigma_{\mu_1\mu_2}$ & $\ell=3$, $P=-1$ & 6.180305(18) & 6.26160085 & 1.315\% \\ 
$\partial_{\nu_1} \partial_{\mu_1} \sigma_{\mu_1\mu_2}$ & $\ell=2$ & 6.180305(18) & 6.29114975 & 1.794\% \\ 
$\square \sigma_{\mu_1\mu_2}$ & $\ell=2$ & 6.180305(18) & 6.39595149 & 3.489\% \\ 
$\varepsilon_{\mu_2 \rho \tau}\partial_{\rho} \partial_{\mu_1} \sigma_{\mu_1\mu_2}$ & $\ell=1$, $P=-1$ & 6.180305(18) & 6.42999132 & 4.040\% \\ 
$\partial_{\mu_1}\partial_{\mu_2} \sigma_{\mu_1\mu_2}$ & $\ell=0$ & 6.180305(18) & 6.52321841 & 5.548\%
\\ \hline\hline
\end{tabular} \end{table}

\begin{table}[ht]
\setlength{\tabcolsep}{0.2cm}
\renewcommand{\arraystretch}{1.4}
    \centering
    \caption{Conformal multiplet of $\sigma_{\mu_1\mu_2\mu_3}$. \label{tab:sigmaL3}}
    \begin{tabular}{cccccc} \hline\hline
Operator & Quantum number  & CB data & $N=16$& Errors \\
$\sigma_{\mu_1\mu_2\mu_3}$ & $\ell=3$ & 4.63804(88) & 4.60892045 & 0.628\% \\ $\partial_{\nu_1}\sigma_{\mu_1\mu_2\mu_3}$ & $\ell=4$ &  5.63804(88) & 5.56345584 & 1.323\% \\ 
$\varepsilon_{\mu_3 \rho \tau}\partial_{\rho} \sigma_{\mu_1\mu_2\mu_3}$ & $\ell=3$, $P=-1$ & 5.63804(88) & 5.6704459 & 0.575\% \\
$\partial_{\mu_1}\sigma_{\mu_1\mu_2\mu_3}$ & $\ell=2$ & 5.63804(88) & 5.79746571 & 2.828\% \\
$\square\sigma_{\mu_1\mu_2\mu_3}$ & $\ell=3$ & 6.63804(88) & 6.74065848  & 1.546\% \\ 
$\partial_{\nu_1}\partial_{\mu_1}\sigma_{\mu_1\mu_2\mu_3}$ & $\ell=3$ & 6.63804(88) & 6.88182226 & 3.672\% \\
$\varepsilon_{\mu_3 \rho \tau}\partial_{\rho} \partial_{\nu_1}\sigma_{\mu_1\mu_2\mu_3}$& $\ell=4$, $P=-1$ & 6.63804(88) & 6.57417625 & 0.962\% \\
$\varepsilon_{\mu_3 \rho \tau}\partial_{\rho} \partial_{\mu_1}\sigma_{\mu_1\mu_2\mu_3}$& $\ell=2$, $P=-1$ & 6.63804(88) & 6.93133276 & 4.418\% \\ 
$\partial_{\mu_1} \partial_{\mu_2}\sigma_{\mu_1\mu_2\mu_3}$& $\ell=1$ & 6.63804(88) & 6.9490099 & 4.685\%
\\ \hline\hline
\end{tabular} \end{table}

Finally, as we discussed in the main text, one of the most surprising aspects  of the fuzzy sphere scheme is that the IR CFT emerges in incredibly small system sizes. 
The best illustration is the observation that our model with only $N=4$ spins (electrons) (Tab.~\ref{tab:N4}) already produces 6 primaries and the approximate conformal invariance.
All  calculations can be done on the laptop, where the $N=16$ spins requires around 16G memory, and the computation took around 30 minutes on a M1 Macbook.

\begin{table}
\setlength{\tabcolsep}{0.2cm}
\renewcommand{\arraystretch}{1.4}
    \centering
    \caption{The rescaled energy gaps of all states of the fuzzy sphere model with $N=4$ spins (electrons). All the states seem to be the Ising CFT states with a small finite size corrections for most of them. \label{tab:N4} }
\begin{tabular}{ccccccccccc} \hline\hline 
& $\sigma$ & $\partial_{\mu_1}\sigma$ & $\square\sigma$ & $\partial_{\mu_1}\partial_{\mu_2}\sigma$ & $\partial_{\mu_1}\partial_{\mu_2}\partial_{\mu_3}\sigma$ & $\partial_{\mu_1}\square \sigma$ & $\sigma_{\mu_1\mu_2}$ & $\sigma_{\mu_1\mu_2\mu_3}$ \\
Bootstrap & 0.518 &  1.518 &  2.518 &  2.518 &  3.518 &  3.518 & 4.180 & 4.638 \\ 
$N=4$ & 0.530 & 1.522 & 2.427 & 2.428 & 2.847  & 3.291 & 4.241 & 4.618 \\
Errors & $2.3\%$ & $0.3\%$ & $3.6\%$ & $3.6\%$  & $20\%$  & $6.5\%$ & $1.5\%$ & $0.4\%$  \\ \hline 
& $\epsilon$ & $\partial_{\mu_1}\epsilon$  & $T_{\mu_1\mu_2}$ & $\partial_{\mu_1}\partial_{\mu_2}\epsilon$ & $\square \epsilon$ & $\epsilon'$ & $\partial_{\mu_3} T_{\mu_1\mu_2}$ & $\varepsilon_{\mu_2\rho\tau}\partial_{\rho} T_{\mu_1\mu_2}$  & $\partial_{\mu_3}\partial_{\mu_4} T_{\mu_1\mu_2}$
\\
Bootstrap & 1.413 & 2.413 & 3 & 3.413 & 3.413 & 3.830 & 4 & 4 &5 \\
$N=4$ & 1.382 & 2.337 & 3 & 3.126 & 3.577 & 4.019 & 3.663 & 4.054 & 4.856 \\
Errors & 2.2\% & 3.1\% & NA & 8.4\% & 4.8\% & 4.9\% & 8.4\% & 1.4\% & 2.9\%
\\ \hline \hline
\end{tabular} \end{table}

\end{widetext}

\end{document}